\documentclass[pre,preprint]{revtex4}

\usepackage{graphicx}
\usepackage{amssymb}
\usepackage{amsmath}
\usepackage{mathrsfs}
\usepackage{subfigure,wrapfig}
\usepackage{enumerate}
\newcounter{tempcounter}

\begin{document}

\title{Multiple bubbles and fingers in a Hele-Shaw channel: complete set of steady solutions}

\author{Giovani L.~Vasconcelos}
\email{giovani.vasconcelos@ufpe.br}

\affiliation{Department of Mathematics, Imperial College London, 180 Queen's Gate, London SW7 2AZ, United Kingdom, and
Departamento de F\'{\i}sica,  Universidade Federal de Pernambuco, 50670-901, Recife, Brazil}

\date{29 May 2015}

\begin{abstract}
Analytical solutions  for  both a finite assembly and a periodic array of bubbles steadily 
moving in a Hele-Shaw channel are presented. The  particular case of multiple fingers penetrating into the channel and moving jointly with an assembly of bubbles is also analysed. The solutions are given by
a conformal mapping from a multiply connected circular domain in an auxiliary complex plane to the fluid region exterior to the bubbles. In all cases the desired mapping  is written explicitly in terms of certain special transcendental functions, known as the secondary Schottky-Klein  prime functions. Taken together, the solutions reported here represent the  complete set  of solutions for steady bubbles and fingers in a horizontal Hele-Shaw channel when surface tension is neglected. All previous  solutions under these assumptions are particular cases of the general solutions reported here. Other possible applications of the formalism described here are also  discussed.

\end{abstract}

\maketitle

\section{Introduction}

Interface dynamics in a Hele-Shaw cell---where the motion of the viscous fluids is confined to the narrow gap between two closely spaced parallel glass plates---is a problem of considerable interest both from a theoretical standpoint as a 
moving free boundary problem and from a practical perspective in view of its connection to other important physical systems, such as flows in porous media, dendritic crystal growth and directional solidification \citep{Pelce}. The Hele-Shaw system is particularly interesting when one fluid is much less viscous than the other and  surface tension effects are neglected, for in this case the problem becomes quite tractable mathematically and many steady and time-dependent  solutions have been found since the pioneering work by \citet{ST}, \citet{Saffman1959} and \citet{TS}. 
Recently, deep mathematical connections have  been discovered between Hele-Shaw flows and other problems in mathematical physics, such as integrable systems, random matrix theory and quantum gravity \citep{Mineev}. There is also a close relation between interface dynamics in a Hele-Shaw cell and an important  growth model known as  Loewner evolution \citep{Hohlov,GS,Zabrodin,Duran}. Motivated by these findings, interest in Hele-Shaw flows (or Laplacian growth, as it is also known) has  grown well beyond its original hydrodynamics context and there is now an extensive literature on the subject and its mathematical ramifications; for an overview, see e.g.~the recent monograph by \citet{Vasiliev}.

Yet, despite all these developments, the Hele-Shaw system continues to surprise and reveal more mathematical structures underlying its dynamics. 
Recent investigations  by \citet{VMC2014}---motivated in part by the problem of interface dynamics in a Hele-Shaw cell---led to the discovery of a new class of special functions  (called the secondary Schottky-Klein prime functions) associated with planar multiply connected  domains. These functions are particularly  useful in tackling potential-theory problems   involving multiply connected domains with {\it mixed boundary conditions}.
 One such problem---and the main theme of the present paper---is the motion of  bubbles  
in a Hele-Shaw channel, in which case  the velocity potential satisfies Dirichlet boundary conditions on the bubble interfaces  and  Neumann conditions on the channel walls.

In this paper, the formalism of the secondary prime functions is used to construct exact solutions for
the problem of multiple bubbles steadily translating in a Hele-Shaw channel,  both for a finite assembly of  bubbles and for a periodic stream of bubbles with an arbitrary number of bubbles per period cell. The problem of multiple fingers penetrating into the channel and moving together with an assembly of bubbles is also analysed as a particular case of the multi-bubble solutions (when some of the bubbles become infinitely elongated). In all cases, the solutions are given in terms of a conformal mapping $z(\zeta)$ from a  multiply connected circular domain in an auxiliary complex $\zeta$-plane to the flow region exterior to the bubbles in the  complex $z$-plane.  This  mapping is written as the sum of two analytic functions---corresponding to the complex potentials in the laboratory and co-moving frames---that map the circular domain onto  slit  domains. Analytical formulae for these slit maps  are obtained in terms of 
 the secondary Schottky-Klein (S-K) prime functions, which then allows us to obtain an explicit solution for the desired mapping $z(\zeta)$.  
 
In the case of a finite assembly of bubbles, a generalised method of images is used  at first to construct the relevant complex potentials and then the resulting expressions (containing an infinite product of terms) are  recast in terms of the secondary prime functions. This function theoretic formulation is more advantageous in that not only does it have
a firmer mathematical basis (the theory of compact Riemann surfaces and their associated prime functions), 
but it can also handle more general cases, such as periodic solutions,  that are not easily tackled by the heuristic method of images.  

Solutions for multiple steady bubbles in a Hele-Shaw channel were first obtained  by the author \citep{GLV1994,GLV2001} for the  cases when the bubbles either are symmetrical about the channel centreline or have fore-and-aft symmetry. In such cases, the fluid region can be reduced by virtue of symmetry to a simply connected domain, whereby the desired mappings can be constructed via the Schwarz-Christoffel  formula. Solutions for an arbitrary number of steady bubbles in an {\it unbounded} cell were obtained by \cite{Crowdy2009a} in terms of the (primary) Schottky-Klein prime function. \citet{Crowdy2009b} also considered the case of a finite assembly of steady bubbles in a channel with no assumed symmetry,  but it was subsequently found that the second family of prime functions used in his approach was not correctly  defined (D. G. Crowdy 2011, personal communication). Exact solutions for this problem were later obtained  by \citet{GV2014} 
using an alternative method based on the generalised Schwarz-Christoffel mapping for multiply connected domains. Their solution is expressed
in the form of an indefinite integral  whose  integrand consists of  products of (primary) S-K prime functions and which contains several accessory parameters that need to be determined numerically. The solutions reported here  for  multiple Hele-Shaw bubbles in a channel are based on an entirely different approach and have the advantage that they are given by an explicit analytical formula in terms of the secondary prime functions, with no accessory  parameters whatsoever. Furthermore, they can be  used to generate new solutions for multiple fingers moving together with an assembly of  bubbles, as will be seen later.

In the case of a periodic array of bubbles, solutions were first obtained by \citet{Burgess} for  a single stream of  symmetrical bubbles.  This class of solutions was later extended to include the case of  multiple  bubbles per period cell  under certain symmetry assumptions \citep{GLV1994,Silva1,Silva2}. The  new family of periodic solutions reported here is much more general in that  it describes an arbitrary stream of {\it groups of bubbles}, with 
no symmetry restriction  on the geometrical arrangement of the bubbles within a period cell. Here again the solutions are given in analytical form in terms of the secondary prime functions,  making the computation of the bubble shapes  a rather simple task, once the preimage domain in the $\zeta$-plane is specified.

In light of the existence of this large class of exact solutions for multiple bubbles, it can be argued that the variety of forms 
observed  by \citet{Maxworthy}, in his experiments  on bubbles rising in an inclined Hele-Shaw cell, is in part  a manifestation of this multitude of  solutions. Further studies would of course be required for a more direct comparison between theory and experiments, but it is worth pointing out that a good agreement was already obtained for the case of 
a small bubble at the nose of a larger bubble \citep{Maxworthy2, GLV2000}.
It is to be noted, however, that not all exact solutions reported here 
are expected to have experimental counterparts,  since they correspond to an idealised model where surface tension and three-dimensional thin-film effects are neglected.

It is also important to emphasise that obtaining analytical solutions for steady Hele-Shaw flows  naturally paves the way to finding time-dependent solutions, as
the form of the steady solutions  often suggests an ansatz for the time-dependent ones \citep[see, e.g.][]{RMP1986}. In this  context, the problem of finding exact solutions for steady Hele-Shaw flows is of physical and mathematical interest 
not only on its own merit but also because it serves as a starting point 
to study 
more general interfacial  problems. 
(The possibility of extending the steady solutions reported herein to the time-dependent case will be briefly discussed towards the end of the paper.)

The analysis presented here for Hele-Shaw bubbles might also find applications in other related problems, such as hollow vortices and streamer discharges in a strong electric field. A hollow vortex is a vortex whose fluid in the interior is at rest (and hence at constant pressure), and so it can be viewed as a bubble with non-zero circulation.
The formalism of the primary S-K prime functions has been used to find solutions for a pair  of translating hollow vortices \citep{hollow2}
as well as for a von K\'arm\'an street  of hollow vortices \citep{hollow1}. It is thus possible that the  present method of analysis, involving multiple bubbles, may be adapted  to study more general configurations of hollow vortices. In the case of
steady streamers in strong electric fields \citep{streamer1}, the governing equations are identical to those for  Hele-Shaw flows, with a streamer corresponding to a bubble or finger, and so the solutions  presented here are likely to be relevant for the problem of multiple interacting streamers \citep{streamer2}.

The paper is organized as follows. In \S\ref{sec:PF},  the problem of an assembly of a finite number of  steady bubbles   in a Hele-Shaw channel is formulated in terms of a conformal mapping $z(\zeta)$ from a  circular domain in an auxiliary complex $\zeta$-plane to the fluid region in the physical $z$-plane. The Schottky groups  associated with this circular domain and their corresponding Schottky-Klein prime functions  are discussed in \S\ref{sec:SK}. The formalism of the secondary prime functions  is then used, in \S\ref{sec:GS},  to obtain an analytical solution for the  mapping  $z(\zeta)$. For ease of presentation, here the problem is first solved by the method of images and then the results  are recast  in terms of the prime functions.  Configurations with multiple fingers moving together with a group of bubbles are discussed in \S\ref{sec:MF}, as a particular case of the general solution for  an assembly of bubbles. In \S\ref{sec:PS}, the case of a periodic array of steady bubbles in a Hele-Shaw channel is considered. Here a fully fledged function-theoretic approach is used to construct  an explicit solution for the corresponding mapping $z(\zeta)$ in terms of the secondary prime functions.  We conclude the paper by briefly discussing, in \S\ref{sec:Dis}, the main features of our results as well as  other possible applications of the analysis presented herein.

\section{Finite assembly of bubbles: problem formulation}

\label{sec:PF}

\begin{figure}
\centering 
\subfigure[\label{fig:1a}]{\includegraphics[width=0.45\textwidth]{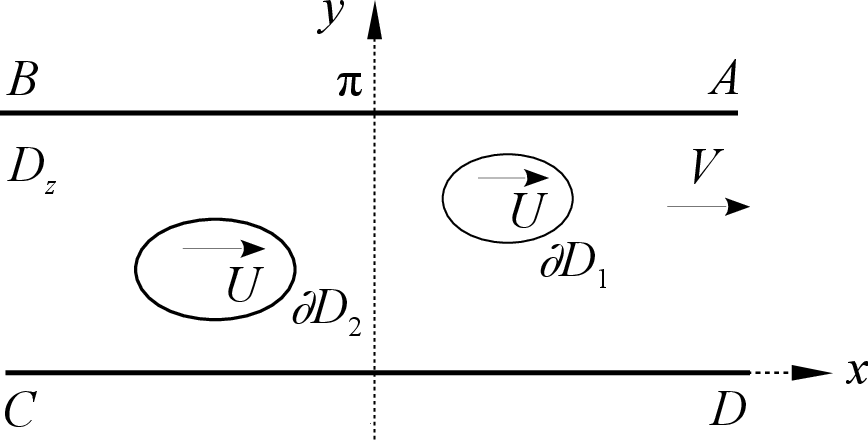}}\qquad
\subfigure[\label{fig:1b}]{\includegraphics[width=0.45\textwidth]{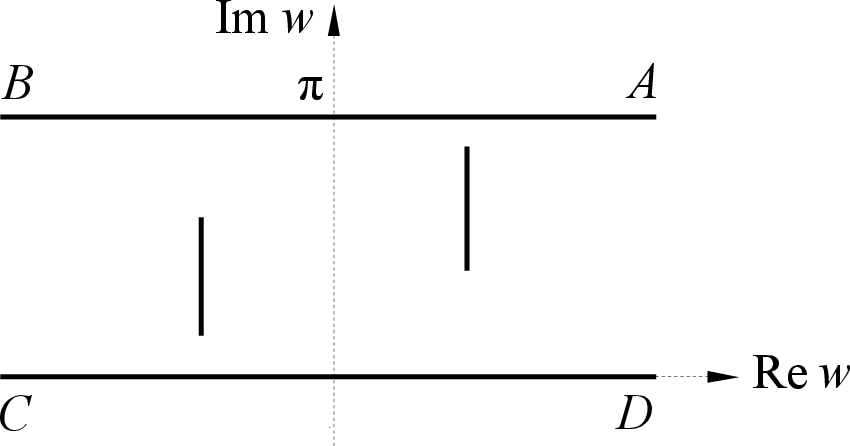}} 
\subfigure[\label{fig:1c}]{\includegraphics[width=0.45\textwidth]{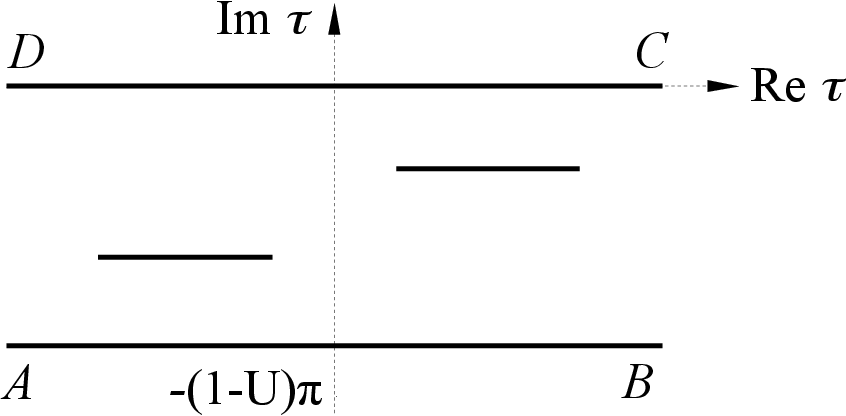}} \qquad
\subfigure[\label{fig:1d}]{\includegraphics[width=0.4\textwidth]{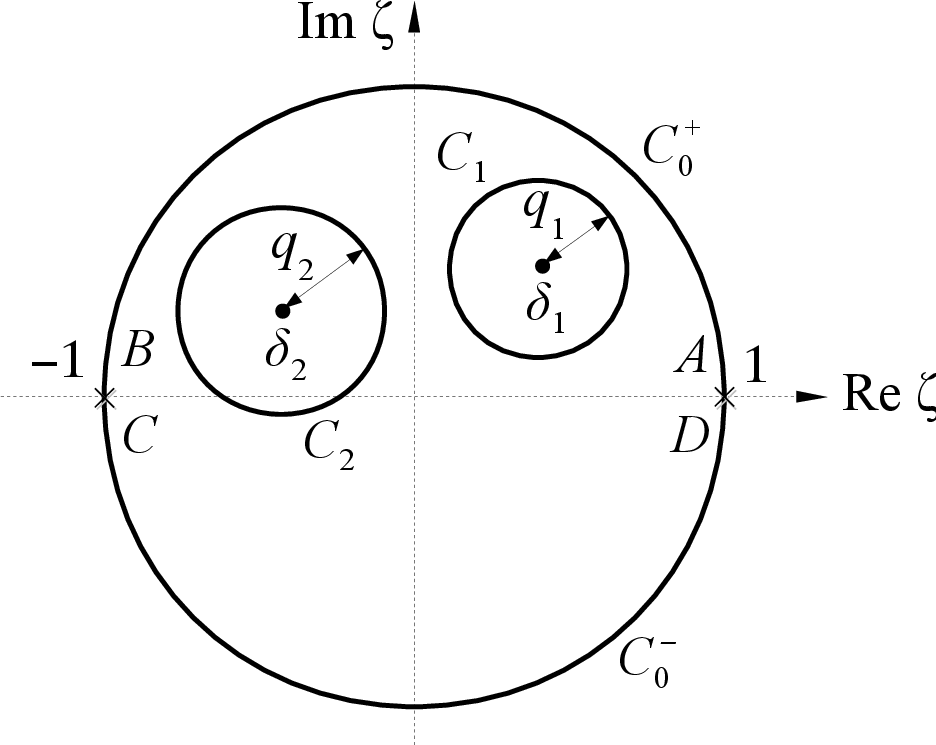}} 
\caption{The flow domains for a finite assembly of bubbles in: (a)  the physical $z$-plane; (b)  the plane of the complex potential in the laboratory frame;  (c)  the plane of the complex potential in the moving frame; and (d)  the auxiliary complex $\zeta$-plane.}
\label{fig:1}
\end{figure}

Here we consider the problem of an assembly of a number $M$ of bubbles of a fluid of negligible viscosity translating  uniformly with speed $U$ parallel to the $x$ axis in a horizontal Hele-Shaw channel filled with a viscous fluid; see figure \ref{fig:1a} for a schematic for the case $M=2$.  The fluid velocity at infinity in front of and behind the bubbles is denoted by $V$, where $U>V$.
 To avoid a proliferation of factors of $\pi$ in our expressions, it is assumed that the channel has width equal to $\pi$; and without loss of generality the far-field velocity is set to unity, i.e.~$V=1$. It is also assumed that the pressure inside each bubble is constant and surface tension effects are neglected, so that the viscous fluid pressure on each bubble boundary is taken to be constant (i.e.~equal to the pressure inside the bubble).

\subsection{The complex potentials}

\label{sec:CP}

As is well known \citep{Howison1992}, the motion of a viscous fluid in a Hele-Shaw cell can be described by a complex potential 
\begin{equation}
F(z,t)=\phi(x,y,t) +\mathrm{i}\psi(x,y,t)
\label{eq:w}
\end{equation}
where  $z=x+\mathrm{i}y$ and the velocity potential $\phi(x,y,t)$ is given by Darcy's law:
\begin{equation}
\phi(x,y,t)=-\frac{b^{2}}{12\mu} p(x,y,t).
\end{equation}
Here $b$ is the cell gap, $\mu$ is the fluid viscosity,   $p(x,y,t)$ is the pressure and  $\psi(x,y,t)$ is the  stream function conjugated to $\phi(x,y,t)$.

Since it is assumed that the bubbles all move with a constant speed $U$, the complex potential $F(z,t)$ has a trivial dependence on the time $t$, namely
\begin{align}
F(z,t)=w(z-Ut),
\end{align}
for some function $w(z)$ to be determined. 
Note that $w(z)$ represents the complex potential in the laboratory frame expressed in terms of the coordinates of a frame of reference co-travelling with the bubbles. It is therefore convenient to introduce a second complex potential, $\tau(z)$, defined by
\begin{equation}
\tau(z)=w(z)-Uz,
\label{eq:tau}
\end{equation}
which describes the flow in the frame of reference  co-moving with the bubbles. Henceforth, it will be understood that the variable $z=x+\mathrm{i}y$ labels points in the co-moving frame.

Now let  $D_z$  denote the fluid region in the $z$-plane exterior to the bubbles, and denote  by $\partial D_j$, for $j=1,...,M$, the boundary of the $j$-th bubble; see figure \ref{fig:1a}. The complex potential $w(z)$ must be analytic in $D_z$ and satisfy the following  boundary conditions:
\begin{equation}
{\rm Im}\left[w(z)\right]=0\qquad\mbox{on}\qquad y=0,
\label{eq:BC1a}
\end{equation}
\begin{equation}
{\rm Im}\left[w(z)\right]=\pi \qquad\mbox{on}\qquad y= \pi,
\label{eq:BC1b}
\end{equation}
\begin{equation}
{\rm Re}\left[w(z)\right]=c_j \qquad \mbox{for}\qquad z\in\partial D_{j},
\label{eq:BC2}
\end{equation}
where  $\{c_j|j=1,...,M\}$ is a set of real constants.
Conditions (\ref{eq:BC1a}) and (\ref{eq:BC1b}) simply state that the channel walls at $y=0$ and $y=\pi$ are  streamlines of the flow, whereas (\ref{eq:BC2})  follows from the fact that the pressure is constant on each bubble boundary (i.e.~the constants $c_j$ are related to the values of the pressure in each bubble).  Furthermore, as the  far-field flow is uniform with unity velocity, it follows that  $w(z)$ has the following asymptotic behaviour at infinity:
\begin{equation}
w(z) \sim z\qquad \mbox{for}\qquad |x|\to\infty.
\label{eq:winfty}
\end{equation}
From (\ref{eq:BC1a})--(\ref{eq:winfty}) one then concludes that the flow domain, $D_w$, in the $w$-plane is a horizontal strip of width $\pi$ containing $M$ vertical slits in its interior, where each slit corresponds to a bubble in the $z$-plane; see figure \ref{fig:1b}. 

Let us now consider the boundary conditions satisfied by $\tau(z)$. As  the bubble boundaries $\partial D_{j}$ are streamlines of the flow in the co-moving frame, one has that
\begin{equation}
{\rm Im}\left[\tau(z)\right]= d_j\qquad \mbox{for}\qquad z\in\partial D_{j},
\label{eq:BC4}
\end{equation}
where  $\{d_j|j=1,...,M\}$ is a set of real constants (corresponding to the values of the streamfunction in the moving frame on each bubble boundary). Furthermore, conditions   (\ref{eq:BC1a}), (\ref{eq:BC1b}) and (\ref{eq:winfty}) imply, in view of (\ref{eq:tau}), that
\begin{equation}
{\rm Im}\left[\tau(z)\right]=0\qquad\mbox{on}\qquad y=0,
\label{eq:BC3a}
\end{equation}
\begin{equation}
{\rm Im}\left[\tau(z)\right]=(1-U)\pi\qquad\mbox{on}\qquad y= \pi,
\label{eq:BC3b}
\end{equation}
and
\begin{equation}
\tau(z) \sim (1-U)z\qquad \mbox{for}\qquad |x|\to\infty.
\label{eq:tauinfty}
\end{equation}
From (\ref{eq:BC4})--(\ref{eq:tauinfty}) one readily sees that the flow domain, $D_\tau$, in the $\tau$-plane is a strip of width $(U-1)\pi$, with $M$ horizontal slits in its interior;  see figure \ref{fig:1c}. 

\subsection{Conformal mapping}

\label{sec:CM}
 
We shall seek a solution for the free boundary  problem defined in \S\ref{sec:CP} in terms of a conformal mapping $z(\zeta)$ from a bounded circular domain $D_\zeta$ in an auxiliary complex $\zeta$-plane to the fluid region $D_z$. To be specific,  let $D_\zeta$ be the domain obtained from the unit disk by excising $M$ non-overlapping smaller disks. A schematic of $D_\zeta$ is shown in figure \ref{fig:1d} for the triply connected case  $M=2$. 

 Label the unit circle by $C_0$ and the $M$ inner circular boundaries by $C_1,...,C_M$; and let $\delta_j$ and $q_j$ denote respectively the centres and radii of the circles $C_j$, $j=0,...,M$. 
(Note that $\delta_0=0$ and $q_0=1$.)  
The mapping  $z(\zeta)$ is chosen such that the unit circle $C_0$ maps to the channel walls, whilst the inner circles $C_1,...,C_M$ map to the bubble boundaries $\partial D_1,...,\partial D_M$. This implies that $z(\zeta)$ will necessarily have two logarithmic singularities,  denoted by $\zeta_1$ and $\zeta_2$,  on the unit circle, which map in the $z$-plane to the  two end points of the channel, $x=\mp \infty$, respectively. By the degrees of freedom afforded by the Riemann-Koebe mapping theorem \citep{Goluzin}, we can set $\zeta_1=-1$ and $\zeta_2=1$. With this choice, the upper unit semicircle, $C_0^+$, maps to the upper channel wall ($y=\pi$) and  the lower unit semicircle, $C_0^-$, maps to the lower wall ($y=0$);  see figure \ref{fig:1}. 

\begin{figure}
\centering 
{\includegraphics[width=0.6\textwidth]{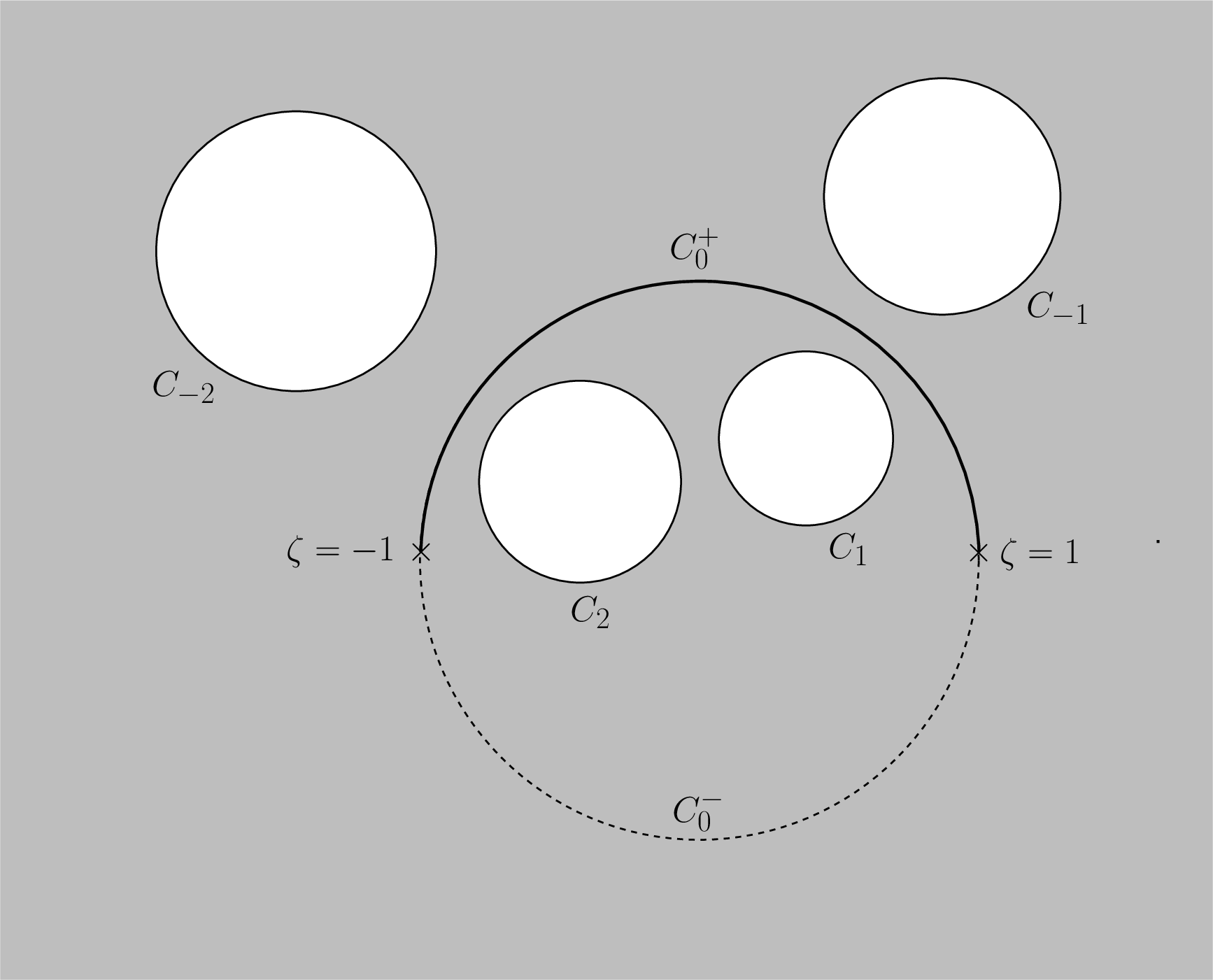}}
\caption{The extended flow domain $F_0$ in the the auxiliary complex $\zeta$-plane.}
\label{fig:F0}
\end{figure}

It is expedient to augment the flow region by reflecting the original channel
 in its lower  wall, thus generating an extended channel defined by $D_z\cup\overline{D_z}$,   where a bar denotes complex conjugation.   (Note that this extended channel contains $2M$ bubbles where each bubble in the lower half-channel is the mirror image of a corresponding bubble in the upper half-channel.)
Accordingly, the extended flow domain, denoted by $F_0$, in the auxiliary $\zeta$-plane is obtained by adding to  
 $D_\zeta$  its reflection in $C_0$:
 \begin{align}
 F_0= D_\zeta \cup \varphi_0(D_\zeta),
 \label{eq:F0}
 \end{align}
 where $\varphi_0(\zeta)=1/\overline{\zeta}$ defines reflection in $C_0$. In addition, a branch cut  must be inserted along $C_0 ^+$ so that the lower and upper sides of this cut map to the upper and lower walls of the extended channel, $y=\pm\pi$, respectively.  
A schematic of $F_0$ is shown in figure \ref{fig:F0}  for the case $M=2$ . 
 
Let us  denote by $C_{-j}$ the reflection of the circle $C_j$ in $C_0$, i.e.~$C_{-j}=\varphi_0(C_{j})$. The region $F_0$ defined in (\ref{eq:F0}) thus corresponds to the exterior of the  circles $C_j$ and $C_{-j}$, for $j=1,...,M$. 
 For convenience of notation, define the following set of labels 
\begin{align*}
K_M\equiv\{-M,...,-1, 1,..., M\},
\end{align*}
 and let $\mathcal{C}_M$ denote the set of  circles bounding $F_0$, that is,
 \begin{align*}
\mathcal{C}_M\equiv\{C_k~|~k\in K_M\}.
\end{align*}
 
Next, introduce the functions $W(\zeta)$ and $T(\zeta)$  through the following compositions:
\begin{align}
W(\zeta)&\equiv w(z(\zeta)) ,\\
 T(\zeta)&\equiv \tau( z(\zeta)).
 \end{align}
The mapping $w=W(\zeta)$ conformally maps $F_0$ onto a slit strip domain in the $w$-plane defined by $D_w\cup\overline{D_w}$. Similarly, the function $\tau=T(\zeta)$ maps $F_0$ onto the slit strip domain $D_\tau\cup\overline{D_\tau}$ in the $\tau$-plane. Note, in particular, that  both $W(\zeta)$ and $T(\zeta)$ must have  logarithmic singularities at $\zeta=\pm 1$. These singularities act as a sink and a source, respectively, for the flows generated by these complex potentials---a fact that will be exploited in \S\ref{sec:GS} to compute explicit formulae for $W(\zeta)$ and $T(\zeta)$ via the method of images. Once these functions  are known, the desired mapping function $z(\zeta)$  that describes the bubble shapes is then given by
\begin{equation}
z(\zeta)=\frac{1}{U}\left[W(\zeta)-T(\zeta)\right],
\label{eq:z1}
\end{equation}
as follows  from (\ref{eq:tau}).

\section{Schottky groups and the Schottky-Klein prime function}

\label{sec:SK}

For any  circular region $F_0$ as defined in (\ref{eq:F0}), one can define a so-called Schottky group generated by the M\"obius transformations that `pair' the circles $C_{-j}$ and $C_j$. Associated with this Schottky group and its subgroups there can  be defined special transcendental  functions, called primary and secondary Schottky-Klein prime functions.
These functions naturally appear in the context of Hele-Shaw flows with multiples bubbles, as will be seen in \S\S\ref{sec:GS}--\ref{sec:PS}, and so  it was thought desirable to present here a brief introduction to the S-K prime functions.

\subsection{The primary S-K prime function}

\label{sec:T0}

Consider the circular domain $F_0$ defined in (\ref{eq:F0}). For $k=-M,...,M$, denote by $\varphi_k(\zeta)$  the 
reflection map in the circle $C_k$ which is defined by
\begin{align}
\varphi_k(\zeta)=\delta_k+\frac{q_k^2}{\overline\zeta - \overline\delta_k}.
\label{eq:varphi}
\end{align}
Now, introduce the following M\"obius maps
\begin{equation}
\theta_k(\zeta)\equiv\varphi_k(1/\overline\zeta)=\delta_k+\frac{q_k^2 \zeta}{1-\overline{\delta}_k\zeta}.
\label{eq:theta}
\end{equation}
Note that $\theta_k(\zeta)$ consists of the composition of a reflection in $C_0$ followed by a reflection in $C_k$, i.e. $\theta_k=\varphi_k\varphi_0$. Alternatively, reflection in $C_k$ can be expressed in terms of $\theta_k$ as 
$\varphi_k=\theta_k\varphi_0$,
or more explicitly:
\begin{align}
\varphi_k(\zeta)=\theta_k(1/\overline\zeta).
\label{eq:vartheta}
\end{align}
 One important property of the maps $\theta_k(\zeta)$ is the following relation:
  \begin{align}
  \theta_{-k}(\zeta)=\theta_{k}^{-1}(\zeta)=\frac{1}{\overline{\theta_k(1/\bar\zeta)}},
 \label{eq:theta_inv}
 \end{align}
where $\theta_k^{-1}(\zeta)$ denotes the inverse of $\theta_k(\zeta)$.  This relation can be derived using geometrical arguments, or it can be verified directly.

Now it may be verified   that $\theta_j(\zeta)$, for $j=1,...,M$,  maps the interior of $C_{-j}$ onto the exterior of   $C_j$. Conversely, the  map $\theta_{-j}(\zeta)$ maps the exterior of $C_{j}$ onto the interior of $C_{-j}$. 
The set $\Theta_0$ consisting of all compositions of the maps $\theta_k(\zeta)$, $k\in K_M$,  defines what is called a classical Schottky group.  The region $F_0$ 
(which we recall consists of  the exterior of the circles ${\cal C}_M$) is called a {\it fundamental region} of the group $\Theta_0$ and 
the maps $\{\theta_j |j=1,..,M\}$ are called  the {\it fundamental generators} of the group. 
For any given Schottky group $\Theta_0$ and  fundamental region $F_0$, the S-K prime function, $\omega(\zeta,\alpha)$, can be defined  for any two points $\zeta, \alpha \in F_0$. 

The S-K prime function admits the following infinite product representation \citep{Baker}:
\begin{equation}
\omega(\zeta,\alpha)=(\zeta-\alpha)\prod_{\theta \in \Theta_0''} \frac{(\zeta-\theta(\alpha))(\alpha-\theta(\zeta))}{(\zeta-\theta(\zeta))(\alpha-\theta(\alpha))}.
\label{eq:SK}
\end{equation}
where $\Theta_0''\subset \Theta_0$ is the set such that for all $\theta\in\Theta_0$, excluding the identity, either  $\theta$ or $\theta^{-1}$ (but not both) is contained in $\Theta_0''$.  For example,  if $ \theta_1\theta_{-2}$ is included in $\Theta_0''$, then $\theta_2\theta_{-1}$ must be excluded.

For later use, it is convenient to quote here the following relation:
\begin{align}
\frac{\omega(\zeta,\alpha)}{\omega(\zeta,\gamma)}=C(\alpha,\gamma) \prod_{\theta \in \Theta_0} \frac{\zeta-\theta(\alpha)}{\zeta-\theta(\gamma)},
\label{eq:SKratio}
\end{align}
where the prefactor  $C(\alpha,\gamma)$  depends on the values of $\alpha$ and $\gamma$ (but not on the point $\zeta$). Note in particular that  the product  in (\ref{eq:SKratio}) is  over the entire group $\Theta_0$. For a derivation of this formula, see, e.g.~\citet{DeLillo2010}; it is also implied by an alternative representation of $\omega(\zeta,\alpha)$ given by \citet{Marshall2005}.

The S-K prime function is intimately connected with the theory of compact Riemann surfaces  \citep{Fay}, but for the present purposes  it suffices to think of it as a special computable function. In this context, a few remarks on the numerical computation of the S-K prime functions are in order.
 The infinite products (\ref{eq:SK}) and (\ref{eq:SKratio}) are known to converge if the circular boundaries of $D_\zeta$ are either centred on the real axis or are sufficiently well separated \citep{Baker, DeLillo2010}. 
 But even when this product formula converges it may be  inefficient
for numerical computation. An alternative
numerical scheme that relies on a representation of the
prime function in terms of much more rapidly convergent Fourier-Laurent sums has been  developed by \citet{CM2007}. This scheme (when applicable) will be used in our numerical computations; see below.

\subsection{Secondary S-K prime functions}

\label{sec:SG}

Given a Schottky group $\Theta_0$ as defined above,   a  family of  Schottky subgroups $\Theta_N\subset \Theta_0$, for $N=1,...,M$,  can be  defined, 
and   prime functions can naturally be associated to them. These so-called secondary prime functions were introduced by \citet{VMC2014}, as building blocks for constructing conformal  mappings for  mixed slit domains, and  are briefly reviewed here.

For a given $N$, with $1\le N\le M$, define $\Theta_{N}$ to be the set of all elements $\theta\in\Theta_0$ such that $\theta$ contains only combinations of an {\it even} number of the maps $\theta_k$, $k=-N,...,N$, $k\ne 0$, but  may contain any number (even or odd) of the other maps $\theta_{k}$, i.e.~for $N+1\le |k|\le M$. For example, for the case $M=3$ and $N=2$, 
one can show that  the group $\Theta_{N}$ is generated by the following maps: $\{\theta_3, \theta_1^2, \theta_1\theta_2, \theta_1\theta_{-2}, \theta_1\theta_3\theta_{-1}\}$; it can indeed be verified that these maps and their inverses generate only (and all) combinations of an even number of the maps $\theta_{\pm1}$ and $\theta_{\pm2}$, but any number of the maps $\theta_{\pm 3}$ can appear.

It may be verified  that the set $\Theta_N$ defined above is itself a Schottky group, which is obviously a subgroup of the original group $\Theta_0$. Associated with the group $\Theta_N$ one can  define a corresponding prime function. To avoid confusion with the primary S-K prime function $\omega(\zeta,\alpha)$ introduced in \S\ref{sec:T} (and associated with the original group $\Theta_0$),  this  {\it secondary} S-K prime function associated with $\Theta_N$ is denoted by $\Omega_N(\zeta,\alpha)$. This function  admits a product representation as in (\ref{eq:SK}), with  the only difference that the product is  over the set $\Theta_N''$ whose definition mirrors that of the set $\Theta_0''$. 

Now fix an integer $l$ such that $1\le l \le N$.
From the definition of $\Theta_N$, one can verify 
that any element $\theta$ of the original group $\Theta_0$ is either an element of  $\Theta_N$  or a composite map of the form $\psi\circ\theta_l$, for some $\psi\in\Theta_N$.
For example,  for the case $M=3$, $N=2$ and $l=1$, we have that $\theta_2\notin \Theta_N$, but  $
\theta_2=\theta_2(\theta_{-1}\theta_1)=(\theta_2\theta_{-1})\theta_1
= \psi\theta_1$, where $\psi=\theta_2\theta_{-1}\in \Theta_N$.
Using this decomposition property of the group $\Theta_0$,  together with  identity (\ref{eq:SKratio}),
one can establish the following relation between the primary and secondary prime functions:
\begin{align}
\frac{\omega(\zeta,\alpha)}{\omega(\zeta,\gamma)}=\tilde{C}(\alpha,\gamma) \frac{\Omega_N(\zeta,\alpha)\Omega_N(\zeta,\theta_l(\alpha))}{\Omega_N(\zeta,\gamma)\Omega_N(\zeta,\theta_l(\gamma))},
\label{eq:SKratio2}
\end{align}
where the prefactor $\tilde{C}(\alpha,\gamma)$ depends only on $\alpha$ and $\gamma$. (This relation holds irrespective of the  choice of $l$, for $1\le l \le N$.)

Note, in particular, that for $N=M$ the subgroup $\Theta_M$ consists of all {\it even} combinations of the maps $\theta_k$, $k\in K_M$; here the generators of the group are the M\"obius maps $\{\theta_1\theta_{k} | k\in K_M\}$ (excluding the identity)  which map the interior of the circles $C_{-k}$ onto the exterior of their images in $C_{1}$.
This subgroup and its associated S-K prime function, $\Omega_M(\zeta,\alpha)$,  play a crucial role in constructing solutions for a finite assembly of bubbles in a Hele-Shaw channel, as discussed in \S\ref{sec:GS}. The function $\Omega_{M-1}(\zeta,\alpha)$, on the other hand,   appears in the problem of periodic arrays of bubbles to be discussed in \S\ref{sec:PS}. 


\section{Solutions for a finite assembly of bubbles}
\label{sec:GS}

In this section the formalism of the S-K prime functions is used to construct explicit formulae for the complex potentials $W(\zeta)$ and $T(\zeta)$ introduced in \S\ref{sec:CM}. A crucial step in this task is the computation of the  infinite sets of images (associated with the sink and source at $\zeta=\pm 1$) that are necessary to enforce the appropriate boundary conditions on the circles $\mathcal{C}_M$ bounding the flow region $F_0$ in the $\zeta$-plane. The location of these images can be expressed in terms of the action of one of the groups $\Theta_N$ on the positions of the original source and sink.  The specific group required  (i.e.~the value of $N$) depends on whether  the flow is described  in the laboratory frame or in the co-moving frame. We begin by considering the complex potential $T(\zeta)$ in the co-moving frame.

\subsection{The function $T(\zeta)$}
\label{sec:T}

Recall that  in the co-moving frame each bubble is a streamline of the complex potential $\tau(z)$, thus implying that the circles $\mathcal{C}_M$ are streamlines of the flow generated by $T(\zeta)$ in the $\zeta$-plane. These boundary conditions can be satisfied with a judicious choice of images, as discussed below.

Consider first the source at $\zeta=-1$. Reflection of this  source in each of the $2M$ circles in $\mathcal{C}_M$ yields a set of $2M$ image sources at the positions $\varphi_k=\theta_k(-1)$,  $k\in K_M$; see (\ref{eq:vartheta}). Here we  used the fact that the image of a point source with respect to a streamline circle is a source of the same strength and located at the corresponding reflection point.

Now, for any given image source, $\theta_{k}(-1)$,  for some $k\in K_M$,  its subsequent reflection in a circle $C_{k'}$, $k'\ne k$, yields another source at the point $\varphi_{k'}(\theta_{k}(-1))=\theta_{k'}(1/\overline{\theta_{k}(-1)})=\theta_{k'}(\theta_{-k}(-1))$,
where we used  (\ref{eq:vartheta}) and (\ref{eq:theta_inv}). Generalising this argument, one can show that reflection in the circles $\mathcal{C}_M$ of the first set of images $\{\theta_k(-1)~|~k\in K_M\}$  generates second-level image sources at the points $\{\theta_{k_1}\circ\theta_{k_2}(-1)~|~k_1, k_2\in K_M, ~k_1+k_2\ne 0 \}$. 
More generally, it  may be verifed that after 
$m$ reflections of the original source in $\mathcal{C}_M$, one obtains a set of sources  whose locations are given by $\{\theta_{k_1}{\circ}\cdots\circ\theta_{k_m}(-1)~|~k_j\in K_M,~k_{j}+k_{j+1}\ne 0,~j=1,...,M \}$. Continuing this procedure {\it ad infinitum}, one  obtains an infinite set of image sources located at the following points: $\{\theta(-1)~|~\theta\in\Theta_0\}$.  A similar procedure for the sink at $\zeta=1$ yields a system of image sinks at the points $\{\theta(1)~|~\theta\in\Theta_0\}$.

The velocity potential $T(\zeta)$ produced by the set of sources and sinks computed above is  given by
\begin{align}
T(\zeta)&=(1-U)
\log
\left(
\prod_{\theta\in\Theta_0}\frac{\zeta-\theta(-1)}{\zeta-\theta(1)}
\right),
\label{eq:Tprod}
\end{align}
where the prefactor  was determined from the requirement that the logarithmic singularities of  $T(\zeta)$   at $\zeta=\pm1$ have the appropriate strength. In other words, when going around either one of these singularities (from one side of the cut to the other) the jump in $T(\zeta)$ must equal $\mathrm{i}2\pi(U-1)$, which corresponds to the width of the extended strip domain  in the $\tau$-plane.

Using (\ref{eq:SKratio}), one can  rewrite  (\ref{eq:Tprod}) in terms of the  S-K prime functions:
\begin{align}
T(\zeta)&=(1-U)\log \left(\frac{\omega(\zeta,-1)}{\omega(\zeta, 1)}\right)+b,
\label{eq:TSK}
\end{align}
where $b$ is an immaterial  constant.
Alternative derivations of this formula directly from the properties of the S-K prime function were given by \citet{Crowdy2009b} and \citet{GV2014}. The derivation presented above is arguably more intuitive in that it is based on the well-known method of images.

For later use, it is convenient to make use of (\ref{eq:SKratio2}) and rewrite (\ref{eq:TSK})  in terms of the secondary prime functions $\Omega_{M}(\zeta,\alpha)$:
\begin{align}
T(\zeta)&=(1-U) \log\left(\frac{\Omega_M(\zeta,-1)\Omega_M(\zeta, \theta_l(-1))}{\Omega_M(\zeta, 1)\Omega_M(\zeta, \theta_l(1))}\right)
\label{eq:T},
\end{align}
where $1\le l \le M$ and an overall  additive constant was omitted since its value is not relevant for the flow.

\subsection{The function $W(\zeta)$}
\label{eq:W}

Here we start by recalling that the bubbles' boundaries are equipotentials of the complex potential $w(z)$ in the laboratory frame, and so the circles $\mathcal{C}_M$ must  be equipotentials of the flow described by $W(\zeta)$. 
Using a similar approach as in \S\ref{sec:T}, one can readily compute the  system of images 
required to satisfy these boundary conditions. The only difference to bear in mind is that  the image of a {\it source} with respect to an equipotential circle is a {\it sink}, and vice-versa.

Consider  the source at $\zeta=-1$. 
Its first set of  images   
with respect to  reflections in the circles $\mathcal{C}_M$  
correspond to $2M$ sinks at the positions $\{\theta_k(-1)~|~k\in K_M\}$. A second round of reflections of these sinks in $\mathcal{C}_M$  then yields sources at the points $\{\theta_{k_1}\circ\theta_{k_2}(-1)~|~k_1, k_2\in K_M, k_1+k_2\ne 0 \}$. Upon  further reflections of these sinks in $\mathcal{C}_M$ one gets a third-level set of sources, and so on, where at each successive level of reflection sources generate sinks and vice-versa. 
In other words, after a sequence of an even number of reflections of the original source 
one gets back a source, whereas an odd number of such reflections produces a sink.
The system of images associated with the primary source at $\zeta=-1$ thus consists of the following two infinite sets:

\begin{enumerate}[i)]
\item ~sources at the points $\{\theta(-1)~|~\theta\in\Theta_M\}$;
\item ~sinks at the points $\{\theta\circ\theta_l(-1)~|~\theta\in\Theta_M\}$;
  \setcounter{tempcounter}{\value{enumi}}
\end{enumerate}

\noindent where we recall that $\Theta_M$ is the set of all {\it even} combinations of the maps $\theta_k$, $k\in K_M$,  and $1\le l \le M$ is an arbitrary integer. 

Similarly, associated  with  the  sink at $\zeta=1$ one finds the following system of images:

\setcounter{enumi}{3}
\begin{enumerate}[i)]
  \setcounter{enumi}{\value{tempcounter}}
\item ~sinks  at the points $\{\theta(1)~|~\theta\in\Theta_M\}$;
\item ~sources at the points $\{\theta\circ\theta_l(1)~|~\theta\in\Theta_M\}$.
\end{enumerate}
Note that in writing down the locations of the images in sets ii) and iv) above, use was made of the fact that  that any combination of an {\it odd} number of the maps $\theta_k$, $k\in K_M$, can be written as $\psi\circ\theta_l$ for  some $\psi\in\Theta_M$, as discussed in \S\ref{sec:SG}. 

Given the sets of sources and sinks in i)--iv) above,  it then follows that the resulting complex potential $W(\zeta)$ is
\begin{align}
W(\zeta)=\log\left(\prod_{\theta\in\Theta_M}\frac{(\zeta-\theta(-1) )(\zeta-\theta(\theta_l(1)))}{(\zeta-\theta(1))(\zeta-\theta(\theta_l(-1)))}\right).
\label{eq:Wprod}
\end{align}
where the prefactor (unity) was chosen so that the width of the extended channel in the $w$-plane is equal to $2\pi$. Now using  (\ref{eq:SKratio}), one can  rewrite 
(\ref{eq:Wprod}) in terms of the secondary S-K prime functions $\Omega_M(\zeta,\alpha)$:
\begin{align}
W(\zeta)&=\log \left(\frac{\Omega_M(\zeta,-1)\Omega_M(\zeta, \theta_l(1))}{\Omega_M(\zeta, 1)\Omega_M(\zeta, \theta_l(-1))}\right),
\label{eq:WSK}
\end{align}
where again an irrelevant  additive constant was omitted.
Once the complex potentials $W(\zeta)$ and $T(\zeta)$ have been obtained, the mapping function $z(\zeta)$ immediately follows from  (\ref{eq:z1}), as discussed next.

\subsection{The conformal mapping $z(\zeta)$}
\label{sec:z}

After substituting   (\ref{eq:T}) and (\ref{eq:WSK}) into (\ref{eq:z1}) and performing some simplification, one  finds
\begin{align}
z(\zeta)=\log \left(\frac{\Omega_M(\zeta,-1)}{\Omega_M(\zeta,1)}\right) +\left(1-\frac{2}{U}\right)\log \left(\frac{\Omega_M(\zeta, \theta_l(-1))}{\Omega_M(\zeta, \theta_l(1))}\right).
\label{eq:z2}
\end{align}
 Recall that $1\le l\le M$ is an integer that can be chosen arbitrarily; in specific computations it is convenient to set $l=1$.

 The coordinates ($x_j, y_j)$ of each  bubble interface $\partial D_j$, $j=1,...,M$, are thus given in parametric form by
\begin{equation}
x_j(s)+\mathrm{i}y_j(s)=z(\delta_j+q_j\exp(\mathrm{i}s)), \qquad 0\le s< 2\pi, 
\label{eq:xy}
\end{equation}
with $z(\zeta)$ as in (\ref{eq:z2}).  Note that all the geometrical information about the bubble configuration described by  the solution above is encapsulated in the prescription of the preimage domain $D_\zeta$. This domain is characterised by its $3M$ conformal moduli $\{\delta_j, q_j~|~j=1,...,M\}$, which  correspond physically to the area and centroids of each of the $M$ bubbles. Thus, once the conformal moduli of $D_\zeta$ are prescribed, a solution for a specific  assembly of bubbles is obtained. 

Let us recall here that  solutions with $U=2$ are special in the sense that all other bubble assemblies with different values of $U$ can be obtained from the $U=2$ solutions by a simple rescaling of coordinates (possibly followed by a rigid translation). This fact was first noticed by \citet{millar} in the context of the Taylor-Saffman solution for a single bubble  and later shown to be  also valid for multiple symmetrical bubbles \citep{GLV2001}. In fact, this property holds in general, i.e.~for  an assembly of steady bubbles  with no assumed symmetry, as can be shown by the following argument which parallels that presented by \citet{GLV2001}, albeit with a different notation. 

First note that the complex potential $W(\zeta)$ given in (\ref{eq:WSK}) does not depend explicitly on the velocity $U$, whereas the complex potential $T(\zeta)$  in the moving frame depends on $U$ only via a prefactor; see (\ref{eq:T}). For a given domain $D_\zeta$,  one can then write
\begin{align*}
W_U(\zeta)&=W_2(\zeta), 
\label{eq:WU}\\
T_U(\zeta)&=(U-1)T_2(\zeta),
\end{align*}
where the  subscripts denote the dependence on the velocity $U$. Inserting these equations into (\ref{eq:z1}), one can show after some manipulation that the coordinates $\left(x_j^{(U)}, y_j^{(U)}\right)$ for the $j$-th bubble in an assembly with speed $U$ can be obtained from the coordinates $\left(x_j^{(2)}, y_j^{(2)}\right)$ of the corresponding bubble in the assembly  with $U=2$ by the following  relations:
\begin{align}
x_j^{(U)}(s)=(1+\rho)x_j^{(2)}(s)-\rho c_j, \\
y_j^{(U)}(s)=(1-\rho)x_j^{(2)}(s)-\rho d_j, 
\end{align}
where $ \rho=1-({2}/{U})$. Here $c_j={\rm Re}[W_2(\zeta)]$ and $d_j={\rm Im}[T_2(\zeta)]$ for $\zeta\in C_j$; see (\ref{eq:BC2}) and (\ref{eq:BC4}). One then sees that a bubble with $U>2$ (i.e.~$0<\rho<1$) is obtained from the corresponding shape with $U=2$ by a stretching along the $x$ direction and a contraction along the $y$ direction (followed by a rigid translation); whereas for $1<U<2$ (i.e.~$-1<\rho<0$) the opposite occurs: contraction  along the $x$-axis and stretching along the $y$-axis. For this reason we shall restrict ourselves  to the case $U=2$ when computing specific examples of bubble configurations, as discussed next.

 \subsection{Examples}
 \label{sec:Ex}

\begin{figure}
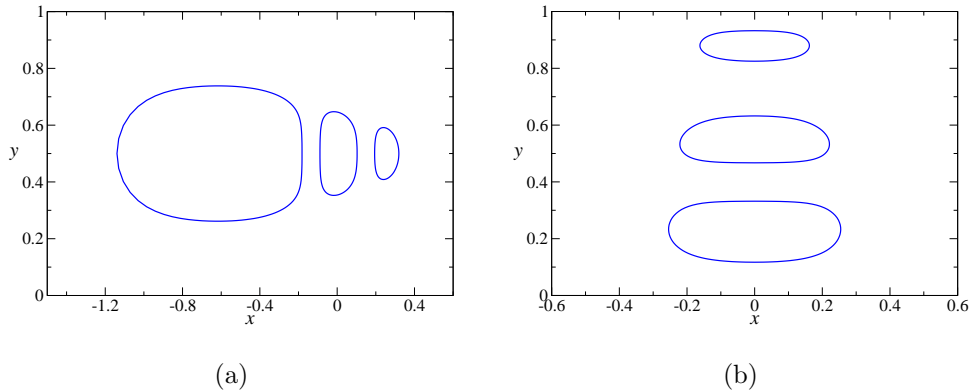

\centering 
\vspace{0.6cm}
\subfigure[\label{fig:3a}]{\includegraphics[scale=0.25]{fig3a.eps}}\qquad
\subfigure[\label{fig:3b}]{\includegraphics[scale=0.25]{fig3b.eps}}
\caption{Examples of assemblies of  three symmetrical bubbles ($M=3$). In  (a)  the bubbles are   symmetric about the centreline, whereas in (b) the bubbles have  fore-and-aft symmetry. The  conformal moduli of $D_\zeta$ are as follows:  $\delta_1=0.4$, $\delta_2=0.0$, $\delta_3=-0.6$, $q_1=0.1$, $q_2=0.1$, $q_3=0.3$ in (a); and $\delta_1=\mathrm{i}0.7$, $\delta_2=\mathrm{i}0.1$, $\delta_3=-\mathrm{i}0.5$, $q_1=q_2=0.3$, $q_3=0.3$ in (b).}
\label{fig:3cent}
\end{figure}

As already mentioned in the Introduction, solutions for multiple steady bubbles in a Hele-Shaw channel were recently  found by \citet{GV2014} in terms of an indefinite integral whose integrand consists of a product of  primary S-K prime functions  and which contains several accessory parameters that need to be determined numerically. The solution (\ref{eq:z2}), in contradistinction,  is expressed as an explicit analytical formula in terms of the secondary prime functions $\Omega_M(\zeta,\alpha)$, with no accessory  parameters. Given a domain $D_\zeta$, the corresponding bubble shapes can be readily obtained upon computation of the relevant prime functions.

In this context, it is important to point out that the numerical scheme developed by \citet{CM2007}  for the computation of the primary S-K function---which avoids the infinite product  and relies on a  more rapidly convergent Laurent series, as discussed in \S\ref{sec:SG}---can be easily adapted for the evaluation of the secondary prime functions $\Omega_M(\zeta,\alpha)$; see  \citet{VMC2014} for  details. 
The numerical computation of  $\Omega_M(\zeta,\alpha)$ can thus be performed in an efficient manner for domains of arbitrary connectivity. 
Using this method, we have reproduced at considerably less computational cost  the specific solutions reported by \citet{GV2014}. Other examples of multi-bubble configurations are discussed below, where  it is assumed that  $U=2$.

In the particular case that the bubbles either are symmetrical about the centreline or have fore-and-aft symmetry, solutions can be obtained by reducing the flow region to a simply connected domain and then applying the standard Schwarz-Christoffel formula \citep{GLV2001}.  These symmetrical solutions  can be recovered from our formula  by simply prescribing a  domain $D_\zeta$ with the appropriate symmetry. More precisely, centreline symmetry is enforced by choosing the centres of all circles $C_j$  on the real axis, whereas bubbles with  fore-and-aft symmetry are obtained by placing the centres of $C_j$ on the imaginary axis. Two examples of assemblies of symmetrical bubbles are shown in figure \ref{fig:3cent}. (In this and in the following figures, we have departed from our original convention and chosen to normalize the channel width to unity for convenience of presentation.)

\begin{figure}
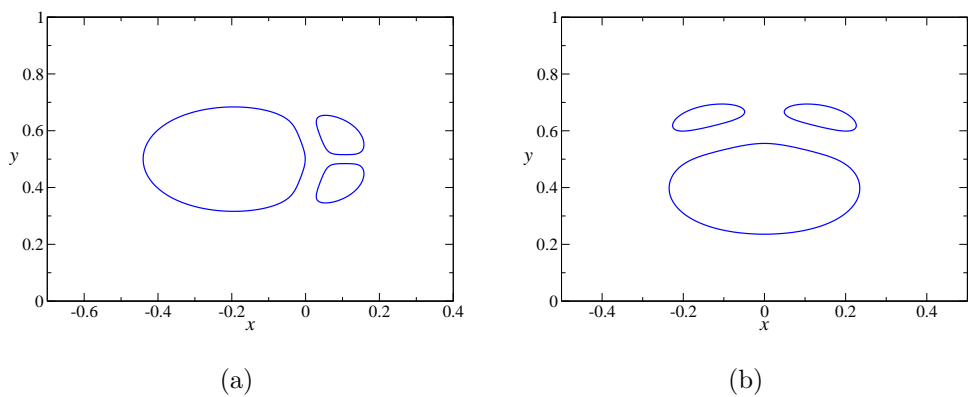

\vspace{0.7cm}
\centering 
\subfigure[\label{fig:4a}]{\includegraphics[scale=0.25]{fig4a.eps}}\qquad
\subfigure[\label{fig:4b}]{\includegraphics[scale=0.25]{fig4b.eps}}
\caption{Examples of assemblies of three bubbles in which the configuration as a whole has either centreline symmetry (a) or fore-and-aft symmetry (b), but where  only one of the bubbles (the largest one in each case) has the overall symmetry of the solution. The  conformal moduli of $D_\zeta$ were chosen as follows: $\delta_1=-0.2$, $\delta_2=0.25+\mathrm{i} 0.12$, $\delta_3=0.25-\mathrm{i} 0.12$, $q_1=0.3$, $q_2=q_3=0.1$ in (a); and $\delta_1=-\mathrm{i}0.2$, $\delta_2=0.2+\mathrm{i} 0.25$, $\delta_3=-0.2+\mathrm{i}0.25$, $q_1=0.3$, $q_2=q_3=0.1$ in (b).}
\label{fig:F0below}
\end{figure}

More generally,  if $D_\zeta$ is reflectionally symmetric about the real (imaginary) axis, then the resulting bubble configuration has centreline (fore-and-aft) symmetry---but not all  bubbles will necessarily have the symmetry of the overall solution (this happens only in the two cases just mentioned).
For instance, in figure \ref{fig:F0below} we show examples of three-bubble assemblies in which the configuration as a whole  has either centreline or fore-and-aft symmetry,  but where only one  of the bubbles (the largest one in each case) possesses the overall symmetry of the solution, whilst the other two bubbles are totally asymmetric.

\begin{figure}
\begin{center}
\vspace{0.8cm}
\centerline{\includegraphics[scale=0.3]{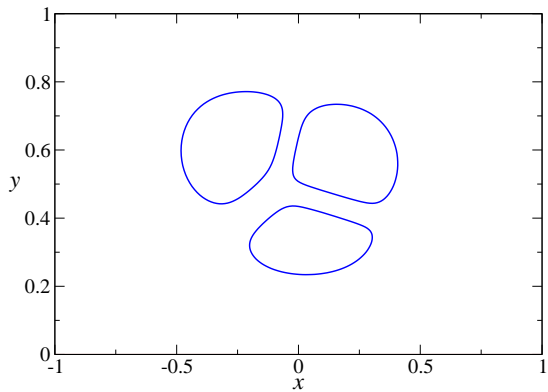}}
\caption{An example of an assembly of three bubbles in a general asymmetric configuration.  Here the conformal moduli of $D_\zeta$ are $\delta_1=0.2+ \mathrm{i}0.2$, $\delta_2=-0.4+ \mathrm{i}0.2$, $\delta_3=- \mathrm{i}0.3$, $q_1=q_2=q_3=0.25$.}
\label{fig:3asym}
\end{center}
\end{figure}

A domain $D_\zeta$ that entails no symmetry yields, of course,  a completely asymmetric bubble configuration. An example of this  case is given in figure \ref{fig:3asym} for an assembly of three asymmetric bubbles.  Solutions for a higher number $M$ of bubbles can  be  handled in a similar manner.

\section{Solutions for multiple fingers and bubbles}

\label{sec:MF}

\begin{figure}
\centering 
\vspace{0.1cm}
{\includegraphics[width=0.6\textwidth]{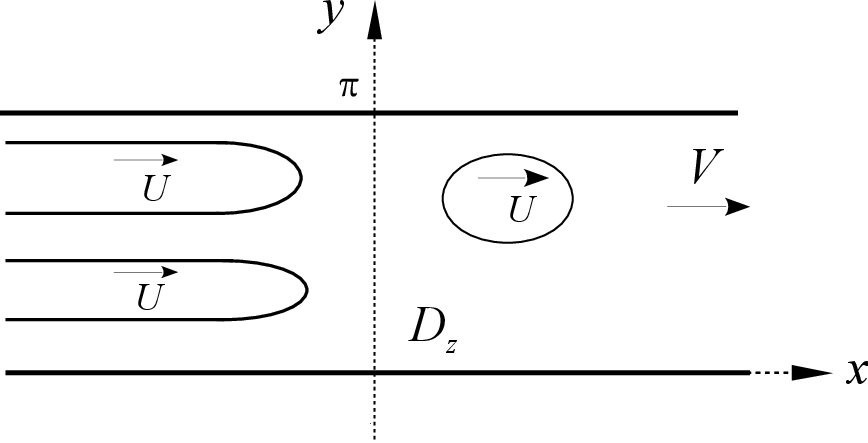}}
\caption{The flow domain  $D_z$ for multiple fingers and bubbles in a Hele-Shaw channel.}
\label{fig:fbz}
\end{figure}

In the instance that some of the bubbles within a multi-bubble solution become infinitely elongated, whilst the other bubbles remain of finite area, one obtains a situation where multiple fingers  penetrate into the channel with an assembly of bubbles moving ahead of the fingers. To be specific, consider a situation with $p$ fingers  and $M$ bubbles. Let us number the fingers from the bottom up and denote by  $\pi\lambda_j$  the width of the $j$-th  finger.  Similarly, let us denote by $\pi\Delta_j$, for $j=1,...,p+1$,  the widths of the fluid gaps separating adjacent fingers or between a finger and a channel wall, with the gaps also numbered from the bottom up.  A schematic of the flow domain $D_z$ in the $z$-plane is shown in figure \ref{fig:fbz}  for the case $p=2$ and $M=1$. 

\begin{figure}
\centering 
{\includegraphics[width=0.6\textwidth]{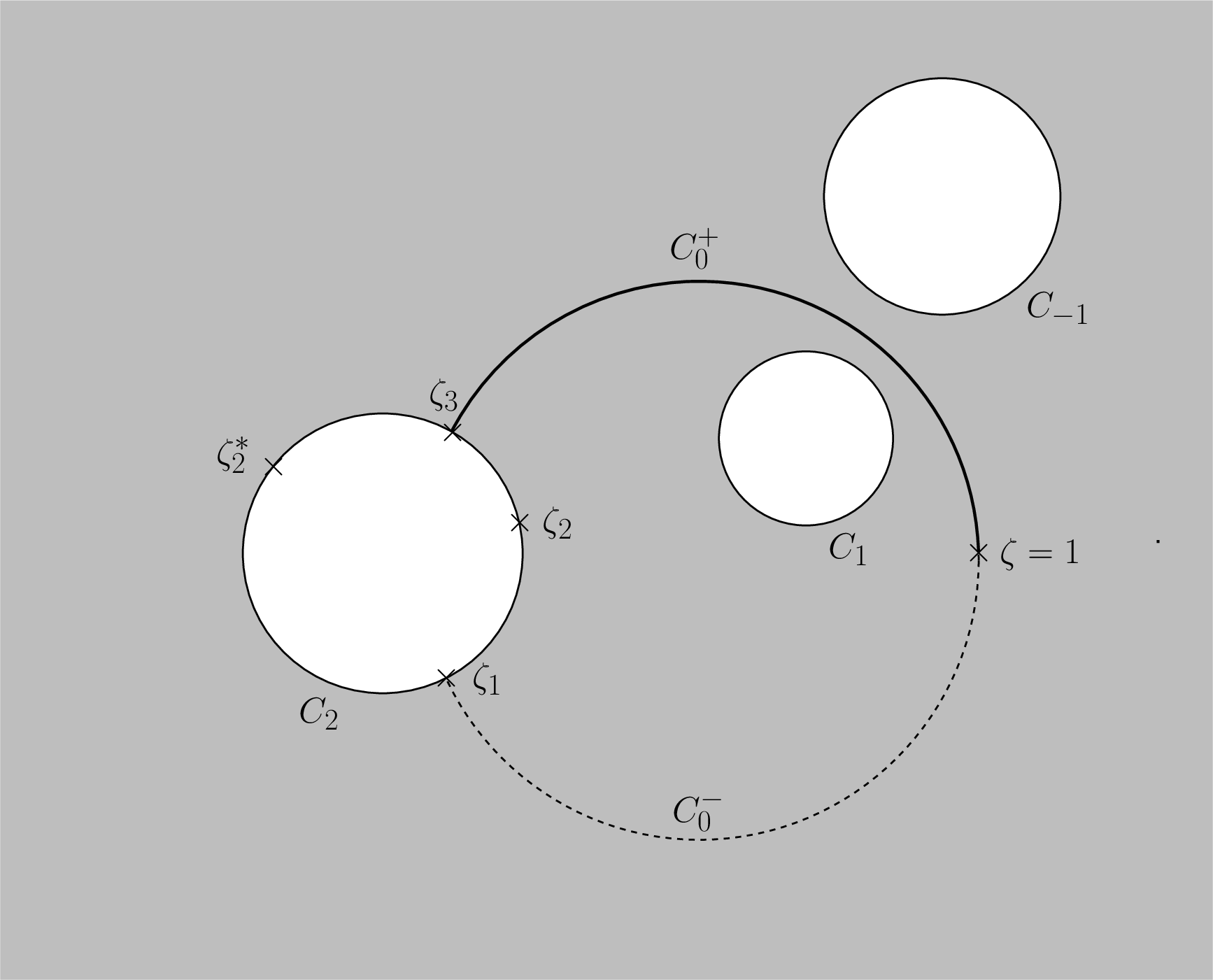}}
\caption{The flow region $\tilde F_0$ in the the auxiliary complex $\zeta$-plane for the case of multiple fingers and bubbles. Here is represented the case of $p=2$ fingers and $M=1$ bubbles in the original channel.}
\label{fig:F0tilde}
\end{figure}

As before, it is  convenient to work in an extended channel, $D_z\cup\overline{D_z}$, containing $2p$ fingers and $2M$ bubbles, where each additional interface is the reflection in the real axis of an interface in the original channel.
The solution to this multifinger problem can be obtained from the multi-bubble solution given in \S \ref{sec:GS} by starting with an assembly of $2(M+p)$ bubbles in the extended channel and then taking the limit in which $2p$ bubbles become infinitely elongated so as to yield the desired  fingers. 
This limit can be easily obtained in the $\zeta$-plane, as follows. The $p$ pairs of circles $C_j$ and  $C_{-j}$, for $j=M+1,...,M+p$,  corresponding to the $2p$ bubbles that will become fingers should coalesce into a single  circle that  {\it orthogonally} intersects  the unit circle $C_0$ and that encloses the point $\zeta=-1$, which was the preimage of $x=-\infty$ in the multi-bubble solution. The other circles $C_k$,  $k\in M$, remain as they are (and so they will map to $2M$ bubbles of finite area). The resulting flow domain,  denoted by $\tilde F_0$, is shown in figure \ref{fig:F0tilde} for the case $M=1$ and $p=2$.

Before proceeding further, let us establish  some notation.  Label by $C_{M+1}$ the circle orthogonal  to $C_0$, and
let $\delta_{M+1}$ and $q_{M+1}$ denote its centre and radius, respectively.  From
 the orthogonality condition one has
\begin{align}
|\delta_{M+1}|^2 = 1 +q_{M+1}^2.
\label{eq:d}
\end{align}
The  M\"obius map $\theta_{M+1}(\zeta)$ associated with reflection in $C_{M+1}$, see (\ref{eq:theta}), is given by
\begin{align}
\theta_{M+1}(\zeta)=\frac{\delta_{M+1} -\zeta}{1-\overline\delta_{M+1}\zeta}.
\end{align}
It  can be verified that this map is of order two: $\theta_{M+1}^2=1$. (Here 1 denotes the identity map.) One may also verify that  $C_{M+1}$ is  invariant under $\theta_{M+1}$ in the following sense:
\begin{align}
C_{M+1}^{\rm(in)}=\theta_{M+1}\left(C_{M+1}^{\rm(out)}\right),
\end{align}
where  $C_{M+1}^{\rm(in)}$  and $C_{M+1}^{\rm(out)}$ denote the segments of $C_{M+1}$ that are inside and outside $C_0$, respectively. 
In fact,  
one can show that $\theta_{M+1}$ maps the interior of $C_{M+1}$ onto its exterior. 
In analogy with the group $\Theta_M$ introduced in \S\ref{sec:SG}, we define the Schottky group $\tilde\Theta_M$ as the set consisting of all even combinations of the maps $\{\theta_{j}| j=1,...,M+1\}$. (The generators of this group are the maps $\{\theta_{M+1}\theta_k~  |~ k\in K_M\}$, which send the interior of $C_{-k}$ onto the exterior of their images in $C_{M+1}$.)

As discussed above, the orthogonal circle $C_{M+1}$ is  to be mapped to the $2p$ fingers, whereas the  circles $C_k$, $k\in K_M$, map to the $2M$ bubbles. This implies, in particular,  that the  function $z(\zeta)$ must  have  $2p$ logarithmic branch points on $C_{M+1}$, corresponding to the left end points of the fingers (i.e.~$x=-\infty$). Let us  denote by $\zeta_j$ the  singularities that lie on $C_{M+1}^{\rm(in)}$ and by $\zeta_j^*$ those lying on $C_{M+1}^{\rm(out)}$, where
\begin{align}
\zeta_j^*\equiv\theta_{M+1}(\zeta_j), \quad j=1,...,p+1.
\end{align}
Note, in particular, that  $\zeta_1$ and $\zeta_{p+1}$ are the points of intersection  between $C_0$ and $C_{M+1}$, and so $\zeta_1^*=\zeta_1$ and $\zeta_{p+1}^*=\zeta_{p+1}$, as these are the two fixed points of the map $\theta_{M+1}(\zeta)$. 
Recall also that by construction the  circle $C_{M+1}$  encloses the point $\zeta=-1$, which was originally mapped to $x=-\infty$. This condition can be fulfilled without loss of generality  by setting $q_{M+1}=1$ and $\delta_{M+1}=-\sqrt{2}$;  this choice will be implied in the remainder of this section.

From the preceding discussion,  it then follows that  solutions with multiple fingers and bubbles can be obtained from the multi-bubble solutions (\ref{eq:z2})  by replacing the  original logarithmic singularity  at $\zeta=-1$  with $2p$ logarithmic singularities at the points $\zeta_j, \zeta_j^*\in C_{M+1}$, for $j=1,...,p+1$. 
In other words, the solution (for $U=2$) given in (\ref{eq:z2})   becomes 
\begin{align}
z(\zeta)&=\log\left(\frac{\prod_{j=1}^{p+1} \left[\tilde \Omega_M(\zeta,\zeta_j)\tilde\Omega_M(\zeta, \zeta_j^*))\right]^{\alpha_j/2}}{\tilde\Omega_M(\zeta, 1)}\right),
\label{eq:zfb}
\end{align}
where  $\tilde \Omega_M(\zeta,\alpha)$ is  the S-K prime function defined over the group $\tilde\Theta_M$ and the parameters $0\le\alpha_j\le 1$ must satisfy the condition 
\begin{align}
\sum_{j=1}^{p+1}\alpha_j= 1,
\label{eq:sumaj}
\end{align}
to ensure single-valuedness of $z(\zeta)$ in $\tilde F_0$.
Note in particular that  the widths of the fluid gaps  (relative to the channel width)  are given by $\Delta_j=\alpha_j/2$, for $j=1,...,p+1$, whereas the  parameters $\zeta_j$, $j=2,...,p$,  control the relative widths of the individual fingers. From (\ref{eq:sumaj}) it  follows that $\sum_{j=1}^{p+1}\Delta_i=\frac{1}{2}$, which implies that   the combined  relative width of the fingers  is  $\sum_{j=1}^p\lambda_j=\frac{1}{2}$, as required by fluid mass conservation (recall that we have set $U=2$).

Once the parameters $\{\alpha_j|j=1,...,p\}$ and  $\{\zeta_j|j=2,...,p\}$ are specified and the conformal moduli $\{\delta_j, q_j | j=1,...,M\}$ of the domain $\tilde D_\zeta$ are prescribed, a specific  solution for $p$ fingers moving together with an assembly of $M$ bubbles in a Hele-Shaw channel is obtained. The shapes of the different interfaces correspond to the images  under the mapping  (\ref{eq:zfb}) of the  circles $C_j$, $j=1,...,M+1$. For example, the coordinates of the $j$-th finger are given by
\begin{align}
x_j^{\rm (f)}(s) + y_j^{\rm (f)}(s) =z\left(-\sqrt{2}+{\rm e}^{is}\right), \quad j=1,...,p,
\label{eq:zfj}
\end{align}
where $\arg(\zeta_j+\sqrt{2})<s<\arg(\zeta_{j+1}+\sqrt{2})$. A similar expression to that in (\ref{eq:xy}) is obtained for the bubble coordinates.

\begin{figure}
\begin{center}
\vspace{0.8cm}
\centerline{\includegraphics[scale=0.3]{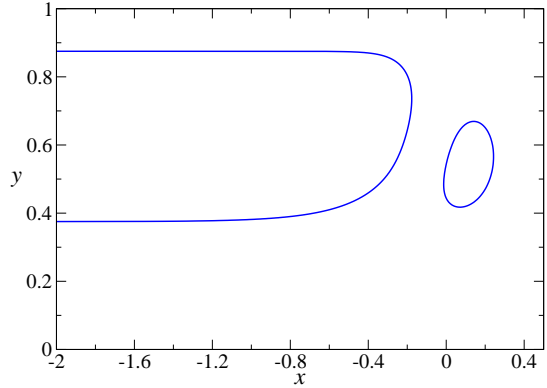}}
\caption{An asymmetric finger with a small asymmetric bubble in front of it. Here  $p=M=1$ and the parameters are $\delta_1=0.2$, $q_1=0.2$, and $\alpha_1=0.75$.}
\label{fig:1fb}
\end{center}
\end{figure}

\begin{figure}
\begin{center}
\vspace{0.5cm}
\centerline{\includegraphics[scale=0.3]{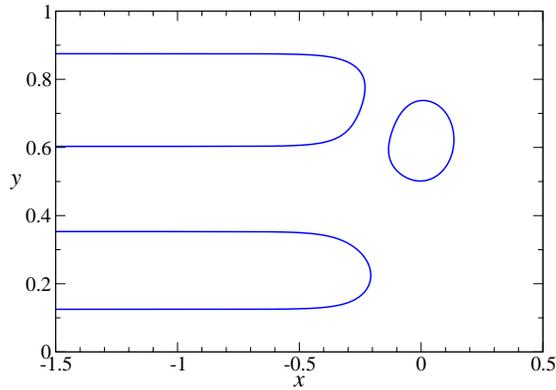}}
\caption{A configuration with  two fingers and one bubble ($p=2$ and $M=1$) corresponding to  the parameters  $\delta_1=\mathrm{i}0.2$, $q_1=0.2$, $\alpha_1=0.25$ and $\alpha_2=0.5$.}
\label{fig:F0fb}
\end{center}
\end{figure}

A related solution for one symmetrical finger with an assembly of symmetrical bubbles in front of it was obtained before  \citep{GLV1999} using Schwarz-Christoffel methods. The present solutions are much more general in that they describe any number of fingers and bubbles with no symmetry assumption and, furthermore, have the advantage of being given by an explicit mapping function from which the shapes of the interfaces can be easily computed.  An example with one asymmetric finger and one asymmetric bubble ($p=M=1$) is shown in figure \ref{fig:1fb}, while  figure \ref{fig:F0fb} shows a configuration with two fingers and one bubble ($p=2$, $M=1$).

Before leaving this section, it is perhaps worth mentioning that in the case when no bubble is present ($M=0$), the S-K prime function becomes a monomial, i.e.~$\tilde\Omega(\zeta,\alpha)=\zeta-\alpha$, yielding a solution (for $p$ fingers only) of the form
\begin{align}
z(\zeta)&=\log\left(\frac{\prod_{j=1}^{p+1} \left[(\zeta-\zeta_j)(\zeta- \zeta_j^*))\right]^{\alpha_j/2}}{\zeta- 1}\right).
\label{eq:zmf}
\end{align}
This solution describes (in a different representation]) the multifinger solutions obtained by \citet{GLV1998}.
Of particular interest is the case of a single finger ($p=1$) for which the expression above reduces to
\begin{align}
z(\zeta)&=-\log \left(\zeta-1\right)+\frac{\alpha}{2}\log\left(\zeta-\zeta_1\right) +\left(1-\frac{\alpha}{2}\right)\log (\zeta- \zeta_2),
\label{eq:zfST}
\end{align}
where $0<\alpha<1$ and $\zeta_2=\overline{\zeta_1}=(-1+i)/\sqrt{2}$.
This recovers (in a different representation)  the asymmetric finger  obtained by \citet{TS} as a generalisation of their previous solution \citep{ST} for a symmetrical finger ($\alpha=1$).

\section{Periodic array of bubbles}
\label{sec:PS}

\begin{figure}
\centering 
{\includegraphics[width=0.6\textwidth]{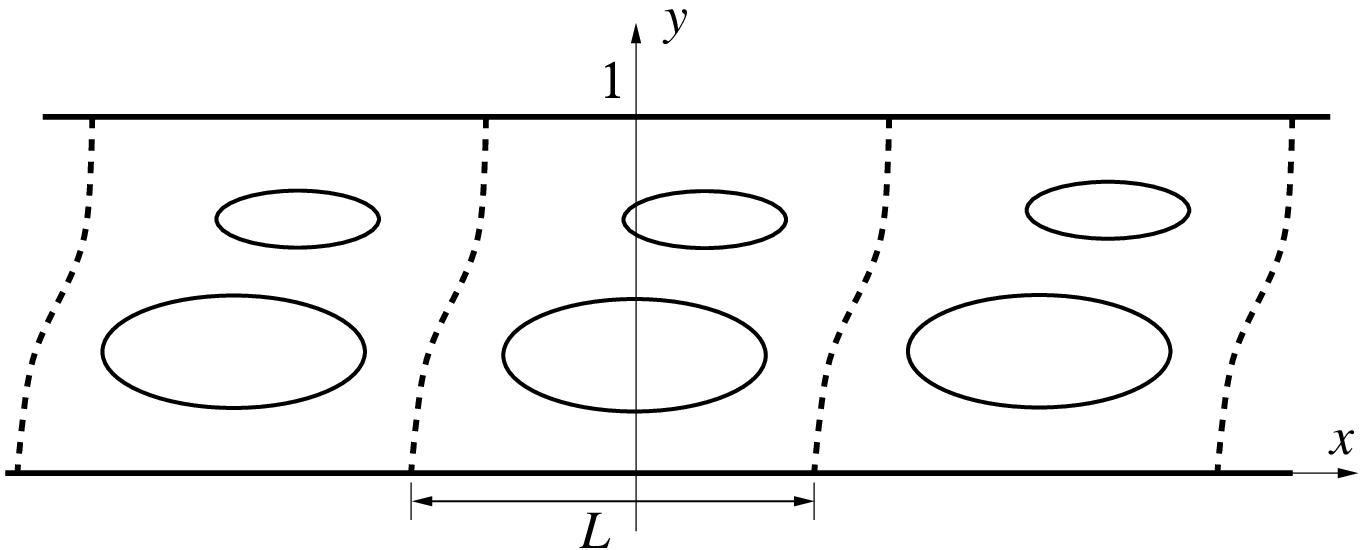}}
\caption{Periodic array of bubbles in a Hele-Shaw chanel.}\label{fig:zp1}
\end{figure}

In this section we consider the case of a periodic array of bubbles  steadily moving in a Hele-Shaw channel. The problem 
is formulated in a general setup where it is supposed that there is an arbitrary number of bubbles per period cell and no assumption is made as to the geometrical arrangement of the bubbles within a period cell; see figure \ref{fig:zp1} for a schematic. This contrasts with all previous periodic solutions \citep{Burgess,GLV1994,Silva1,Silva2} where symmetry requirements are imposed {\it a priori}.  As in the case of a finite assembly of bubbles discussed in \S\ref{sec:GS}, exact solutions for the present problem are obtained in the terms of a conformal mapping from a circular domain to the flow region exterior to the bubbles (within a period cell).
Here,  however, a fully fledged function theoretic approach is used, whereby the desired mapping functions are obtained by directly exploiting the properties of the secondary prime functions.

 \subsection{Problem formulation}

Consider a periodic assembly of  bubbles moving with speed $U$ in a Hele-Shaw channel. Here there are no factors of $\pi$ to worry about and so we set the width of channel to unity.
The average fluid velocity, $V$,  across the channel in the $x$-direction is also normalised to unity, i.e.~$V=1$.
The streamwise period is denoted by $L$, and it is assumed that there are $M-1$ bubbles per period cell. Because of the flow periodicity, the problem can be restricted to the fluid region, $D_z$, within one period cell. A schematic of  $D_z$ is shown in  figure \ref{fig:1ap} for the case $M=3$.

Now let $D_\zeta$ be our usual circular domain with $M$ inner circles; see figure \ref{fig:1bp}.
We shall seek a conformal mapping $z(\zeta)$ from $D_\zeta$ to $D_z$, such that the unit circle $C_0$  maps to the upper edge of the period cell  ($y=1$), the inner circle $C_M$ maps to the lower edge ($y=0$), and the other inner circles  $C_j$, $j=1,...,M-1$,  map to the bubble boundaries $\partial D_j$. In addition, we insert a branch cut from a point on  $C_M$ to a point on  $C_0$ such that the two sides of this cut map to the left and right edges of the period cell, respectively; see figure \ref{fig:1bp}. The specific `path' of the branch cut is not relevant as it  only affects the choice of period cell in the $z$-plane. (In the examples discussed below,  the branch  cut is placed on the positive real axis for convenience.)

 \subsection{General solutions}

\begin{figure}
\centering 
\subfigure[\label{fig:1ap}]{\includegraphics[width=0.4\textwidth]{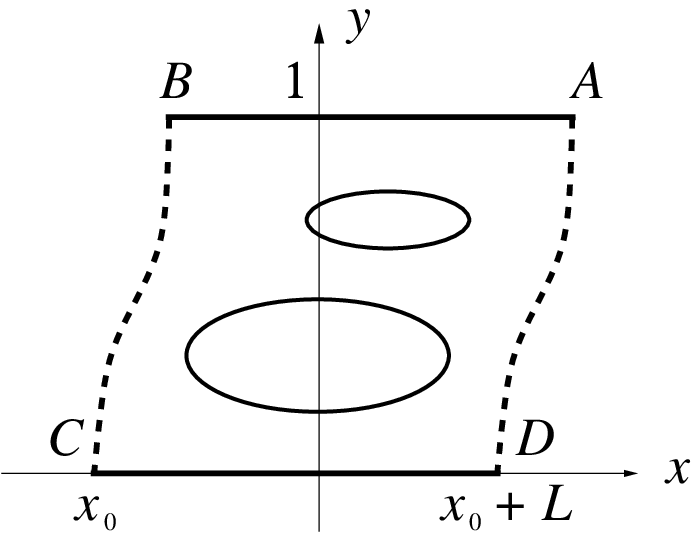}} \qquad
\subfigure[\label{fig:1bp}]{\includegraphics[width=0.4\textwidth]{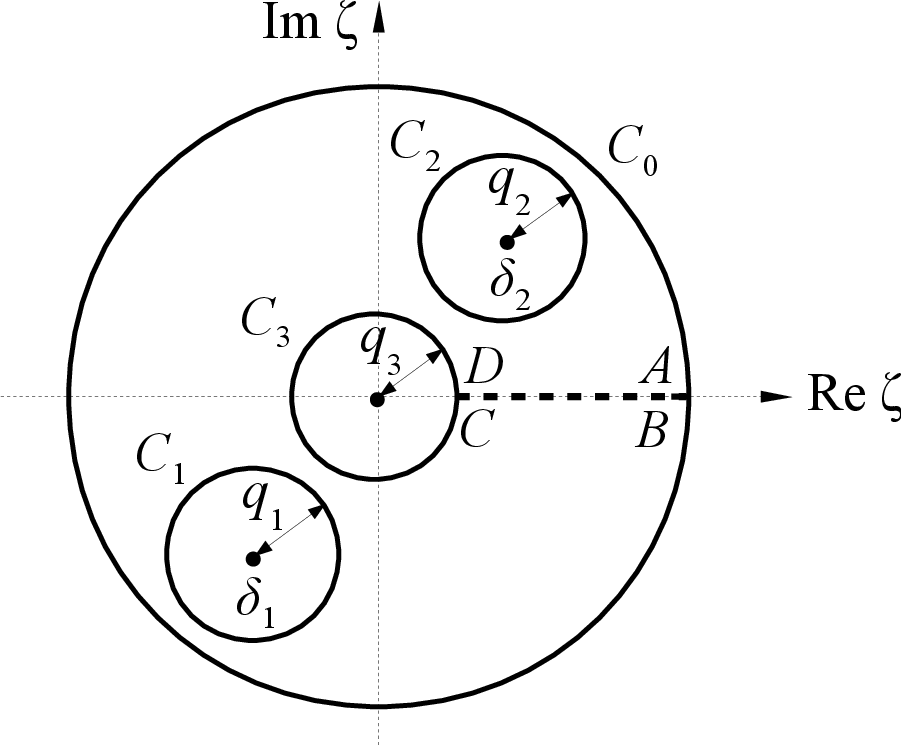}}
\subfigure[\label{fig:1cp}]{\includegraphics[width=0.45\textwidth]{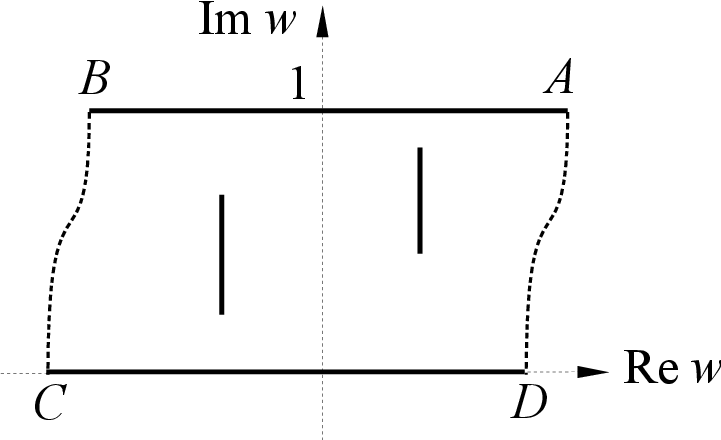}}\quad
\subfigure[\label{fig:1dp}]{\includegraphics[width=0.45\textwidth]{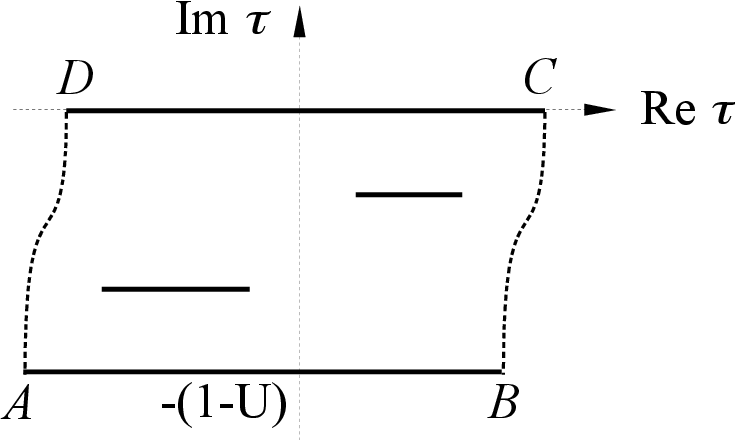}} 
\caption{The flow domain within a period  cell in: (a) the $z$-plane, (b) the  $\zeta$-plane, (c) the $w$-plane, and (d) the $\tau$-plane.}\label{fig:zp}
\end{figure}

As before, let $W(\zeta)$ and $T(\zeta)$  denote the complex potentials in the laboratory and co-moving frames, respectively. Carrying out an analysis analogous to that presented  in \S\ref{sec:CM}, one can show that 
 $w=W(\zeta)$  maps  $D_\zeta$ onto  a  ``rectangular'' region, $D_w$, 
  in the $w$-plane 
 which  is bounded by  two horizontal edges located at ${\rm Im}\, w=0$ and ${\rm Im}\,w=1$ (the images of $C_0$ and $C_M$) and  by two `curved' lateral edges (the image of the branch cut from $C_0$ to $C_M$), and which contains $M-1$ slits  in its interior (the images of $C_j$, $j=1,...,M-1$); see figure \ref{fig:1cp}. In the same vein,  one may verify that $\tau=T(\zeta)$ maps $D_\zeta$ onto a   ``rectangular" domain, $D_\tau$, in the $\tau$-plane
with $M-1$ horizontal slits; 
 see figure \ref{fig:1dp}. 

\begin{figure}
\begin{center}
\subfigure[\label{fig:radial}]{\includegraphics[width=0.4\textwidth]{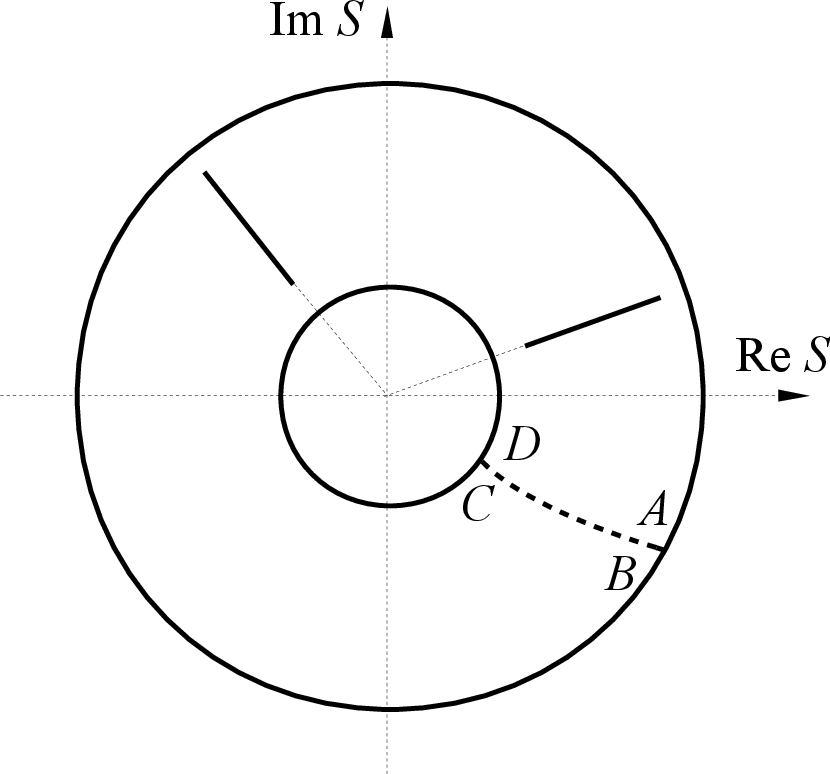}}\qquad\qquad
\subfigure[\label{fig:circ}]{\includegraphics[width=0.4\textwidth]{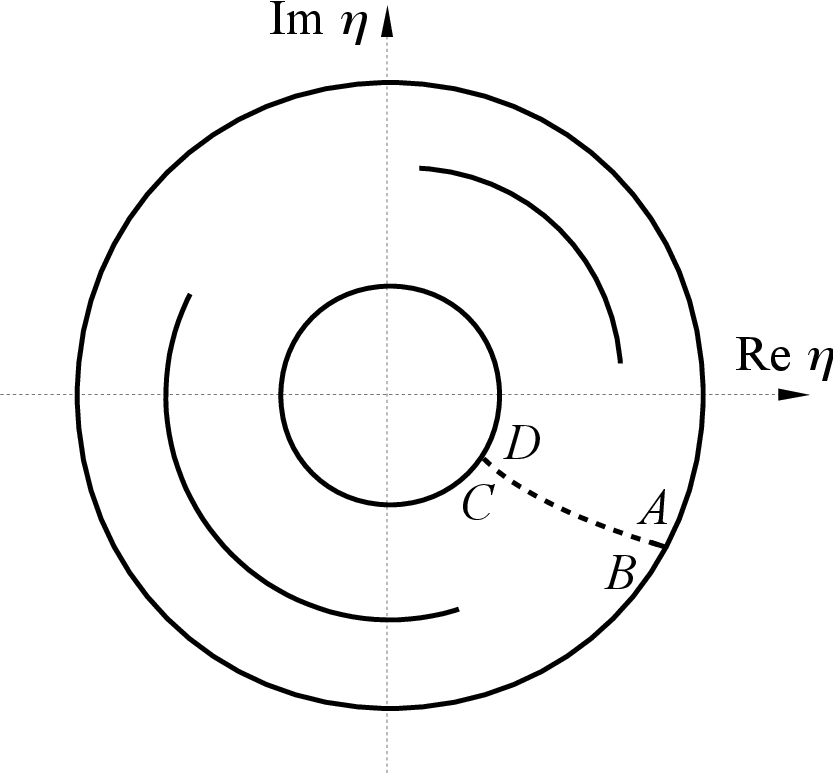}}
\end{center}
\caption{The flow domain in the subsidiary $S$-plane (a) and $\eta$-plane (b).}
\label{fig:annulus}
\end{figure}

To obtain $W(\zeta)$, let us first introduce the following transformation 
\begin{align}
S=\exp\left(-\mathrm{i} \lambda w\right),\qquad \lambda\in\mathbb{R}_{>0},
\label{eq:eW}
\end{align}
which maps  $D_w$ onto a domain in a subsidiary $S$-plane which consists of a concentric annulus with $M-1$ radial slits; see figure \ref{fig:radial}. The real parameter $\lambda$ allows  us the freedom to vary the  modulus of the annulus in the $S$-plane.

Now, it was shown by \cite{VMC2014} that the function 
\begin{equation}
S(\zeta)
=\frac{\Omega_{M-1}(\zeta,\theta_{M}(\alpha))\Omega_{M-1}(\zeta,\theta_l(\alpha))}{\Omega_{M-1}(\zeta,\alpha)\Omega_{M-1}(\zeta,\theta_l(\theta_{M}(\alpha))},
\label{eq:S}
\end{equation}
where $1\le l\le M-1$ and $\alpha\in \varphi_0(D_{\zeta})$,
maps  the circular domain $D_\zeta$ conformally onto a concentric annulus with $M-1$ radial slits. Here $C_0$ is mapped to the outer circumference of the annulus and  $C_M$ maps to the inner circumference, whereas  $C_j$, $j=1,...,M-1$, map to the slits.

Using (\ref{eq:eW}) and (\ref{eq:S}), one finds that the map $w=W(\zeta)$ is of the form
\begin{align}
W(\zeta)&=\mathrm{i} K \log\left[\frac{\Omega_{M-1}(\zeta,\theta_{M}(\alpha))\Omega_{M-1}(\zeta,\theta_l(\alpha))}{\Omega_{M-1}(\zeta,\alpha)\Omega_{M-1}(\zeta,\theta_l(\theta_{M}(\alpha))}\right],\label{eq:Wp}
\end{align}
where  $K$ is a real constant. The value of $K$  is determined from the condition that the ``rectangular" domain $D_w$ has a height equal to $1$, that is,
\begin{align}
{\rm Im}\left[ W(A)-W(D)\right]=1,
\end{align}
where the points $A$ and $D$ correspond to the intersections of the branch cut with the circles $C_0$ and $C_M$, respectively; see figures \ref{fig:1bp} and \ref{fig:1cp}.

Similarly, to obtain the mapping $\tau=T(\zeta)$ one first  applies an exponential transformation
\begin{align}
\eta=\exp\left(\mathrm{i} \lambda \tau \right), \qquad\lambda\in\mathbb{R}_{>0},
\label{eq:eT}
\end{align}
 which maps $D_\tau$ onto a domain $D_\eta$ consisting of a concentric annulus with $M-1$ concentric circular-arc slits; see figure \ref{fig:circ}. Next, we recall that as shown by \citet{CM2007} the  function
\begin{align}
\eta(\zeta)&= \frac{\omega(\zeta,\theta_{M}(\alpha))}{\omega(\zeta,\alpha)},
\label{eq:etaT}
\end{align}
where $\alpha\in \varphi_0(D_\zeta)$,
maps $D_\zeta$ onto  the desired annular slit domain  $D_\eta$. 
(Here $C_0$ maps to the outer circumference of the annulus, $C_M$ maps to the inner circumference, and $C_j$, $j=1,...,M-1$, map to the circular slits.)

Using (\ref{eq:eT}) and (\ref{eq:etaT}), one then finds that
\begin{align}
T(\zeta)&=\mathrm{i} K' \log\left[\frac{\omega(\zeta,\theta_{M}(\alpha))}{\omega(\zeta,\alpha)}\right],
\label{eq:Tp}
\end{align}
where the prefactor $K'$  is determined from the requirement that the height of the domain $D_\tau$ is equal to $U-1$:
\begin{align}
{\rm Im}\left[ T(D)-T(A)\right]=U-1.
\end{align}
In view of (\ref{eq:SKratio2}), one can  rewrite (\ref{eq:Tp}) in terms of the secondary prime functions $\Omega_{M-1}(\zeta,\alpha)$:
\begin{align}
T(\zeta)
=\mathrm{i}K' \log\left[\frac{\Omega_{M-1}(\zeta,\theta_M(\alpha))\Omega_{M-1}(\zeta, \theta_l(\theta_M(\alpha)) }{\Omega_{M-1}(\zeta,\alpha) \Omega_{M-1}(\zeta, \theta_l(\alpha))}\right].
\label{eq:Tp2}
\end{align}

Inserting (\ref{eq:Wp}) and  (\ref{eq:Tp2}) into (\ref{eq:z1}), and performing some straightforward rearrangements, then yields the desired mapping  $z(\zeta)$:
\begin{align}
z(\zeta)
&=\frac{\mathrm{i}K_-}{U}\log\left[\frac{\Omega_{M-1}(\zeta,\theta_M(\alpha))}{\Omega_{M-1}(\zeta,\alpha)}\right]
+ \frac{\mathrm{i}K_+}{U}\log\left[\frac{\Omega_{M-1}(\zeta, \theta_l(\alpha)) }{\Omega_{M-1}(\zeta, \theta_l(\theta_M(\alpha))}\right]
,
\label{eq:zp}
\end{align}
where
\begin{align}
K_\pm=K\pm K'.
\end{align}

\begin{figure}
\begin{center}
\vspace{0.8cm}
\centerline{\includegraphics[scale=0.3]{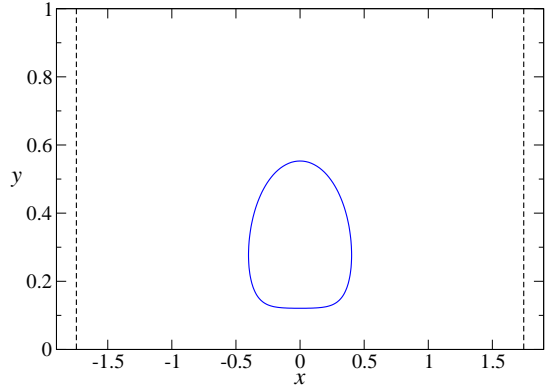}}
\caption{A periodic stream of bubbles with one bubble per period cell ($M=2$). The conformal moduli of $D_\zeta$ are as follows: $\delta_1=-0.6$, $\delta_2=0.0$, $q_1=0.3$, $q_2=0.15$; the corresponding period is  $L=3.49$. The lateral  edges  of the period cell,  $y=\pm L/2$ (dashed lines), are equipotentials of the flow. }
\label{fig:1p}
\end{center}
\end{figure}

Using the degrees of freedom allowed by the Riemann-Koebe mapping theorem \citep{Goluzin}, we can place the centre of the circle $C_M$ 
at the origin, i.e., $\delta_{M}=0$. Its radius, $q_M$, is then a free parameter that essentially controls the period $L$. The remaining $3M-3$ parameters, corresponding to the conformal moduli $\{\delta_j, q_j~|~j=1,...,M-1\}$ of $D_\zeta$, determine the centroid and area of the $M-1$ bubbles in a period cell. Thus, once the domain $D_\zeta$ is prescribed a specific solution for a periodic assembly of bubbles  can be readily computed from (\ref{eq:zp}).

\subsection{Examples}
\label{sec:Exp}

As seen above, obtaining specific solutions for a periodic array of Hele-Shaw bubbles requires computation of  the secondary prime functions $\Omega_{N}(\zeta,\alpha)$ for $N=M-1$. This can be done by using the  infinite product (\ref{eq:SK}) defined over the appropriate group $\Theta_{M-1}$. (Alternative numerical schemes to compute $\Omega_{N}(\zeta,\alpha)$ are  known only for the cases $N=M$ and $N=1$; see \citet{VMC2014}.)  
For the present  purposes, it suffices to truncate the infinite product at the fourth-level maps (i.e.~maps involving up to four generators of the group). Including higher order maps makes no discernible difference within the scale of the figures.

\begin{figure}
\begin{center}
\vspace{0.8cm}
\centerline{\includegraphics[scale=0.3]{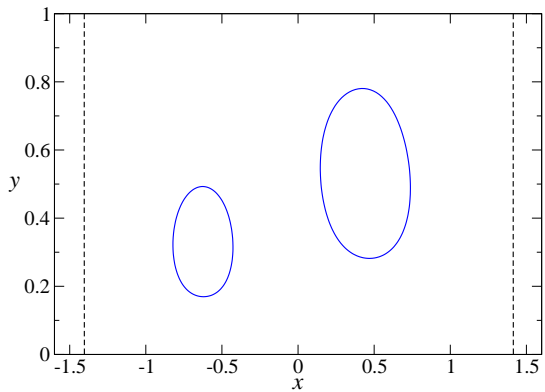}}
\caption{An example of a staggered two-file stream of bubbles. Here $M=3$ and the conformal moduli of $D_\zeta$ are $\delta_1=-0.1+\mathrm{i} 0.5$, $\delta_2=-0.2-\mathrm{i} 0.3$, $\delta_3=0.0$, $q_1=q_2=0.2$, $q_3=0.1$. The period is $L=2.82$ and two period cells are shown.}
\label{fig:F0p}
\end{center}
\end{figure}

An example of a periodic array of bubbles with only one bubble per period cell is shown in figure \ref{fig:1p}. It follows from symmetry considerations  that in this case the bubble shape has fore-and-aft symmetry, so that the vertical lines at $x=\pm L/2$  are  equipotentials of the flow. In particular, it is worth noting that when the bubble is also symmetrical about the centreline, our solution recovers the stream of symmetrical bubbles obtained by  \citet{Burgess}.
As discussed before,  periodic solutions with more general symmetrical arrangements, such as those found by  \citet{Silva1, Silva2}, can easily be reproduced  in our formalism by simply  choosing $D_\zeta$ with the appropriate symmetry. Furthermore,  the analytical solution given in (\ref{eq:zp})  can handle asymmetric configurations with equal ease, as illustrated below. 

\begin{figure}
\begin{center}
\vspace{0.7cm}
\centerline{\includegraphics[scale=0.3]{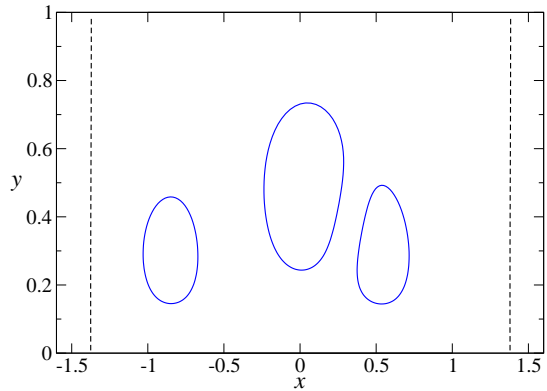}}
\caption{An example of a periodic array of bubbles with three asymmetric bubbles per period cell ($M=4$) corresponding to the  following choice of parameters: $\delta_1=0.2+\mathrm{i} 0.5$, $\delta_2=-0.2-\mathrm{i} 0.5$, $\delta_3=-0.38$, $\delta_4=0.0$, $q_1=q_2=q_3=0.2$, $q_4=0.1$. Here the period is $L=2.76$ and two period cells are shown.}
\label{fig:3p}
\end{center}
\end{figure}

An example of a staggered  two-file array of unequal bubbles is shown in figure \ref{fig:F0p}. For the choice of parameters used in this case (see caption of figure)  the period is $L=2.82$, and two period cells are shown in the figure. Bubble configurations with a higher number of bubbles per period  cell can be computed in a similar manner. An example with three asymmetric bubbles per unit cell is given in figure \ref{fig:3p}, where again two period cells are shown in the figure.

\section{Discussion}

\label{sec:Dis}

The  motion of an assembly of  bubbles in a Hele-Shaw channel is a  free boundary problem which is made  more difficult by the fact that the 
relevant field (the velocity potential)  is defined over a multiple connected domain   and on whose boundaries it satisfies mixed boundary conditions. This means  that in the complex potential plane the flow region is a slit strip domain of mixed type: the bubbles are described by {\it vertical} slits, whilst the channel walls correspond to {\it horizontal} lines. Recently, a formalism based on the secondary S-K prime function was developed to construct 
a large class of such mixed slit maps  \citep{VMC2014}.

Here the  formalism  of the secondary prime functions  was used to compute exact solutions  for multiple bubbles  steadily translating in a Hele-Shaw channel in various configurations: i) finite assembly of bubbles; ii) multiple fingers moving together with an assembly of bubbles; and iii) periodic array of bubbles.  In all cases considered, analytical formulae in terms of the secondary prime functions were obtained for the conformal mapping from a  circular domain to the corresponding flows region in the physical plane.  Several examples of specific solutions for  these distinct arrangements  were given. Taken together,  the results reported here represent the complete set of solutions for multiple steady bubbles and fingers in a horizontal Hele-Shaw channel when surface tension is neglected. 

A variant of the Hele-Shaw problem  that has received less attention is the case when the cell is rotated about the centerline away from the horizontal, which introduces a gravitational potential transverse to the centreline.  To the best of our knowledge, the only known solution for this situation (in the channel geometry) was obtained by \citet{BLT} for the case of a non-symmetric finger. It would be interesting to investigate whether this solution can be extended for the case of multiple bubbles in a rotated channel. The additional complication here is that the flow region in the complex potential plane consists of a strip  with  \emph{slanted} slits (rather than vertical ones), and conformal mappings to this type of slit domains are not yet known. Injection flows in an unbounded Hele-Shaw cell in the presence of a uniform gravitational field have also been considered using Schwarz function-type methods \citep{Entov1995, McDonald2011}. It remains to be seen whether these alternative approaches could be adapted to the channel geometry and  generalised to the multiply connected case.

Another possible extension of the present research would be to consider time evolving bubbles in a Hele-Shaw cell. 
Time-dependent solutions for Hele-Shaw flows date back to the early work by \citet{Saffman1959} who found a solution for a finger growing from a flat interface. Since then,  more general solutions
have been found  both for  the growth of fingers in a channel  \citep{Mineev1994} and for an expanding bubble in an unbounded cell \citep{ShraimanBensimon, BensimonPelce,Howison}. Several exact solutions for doubly connected time dependent Hele-Shaw flows have also been obtained  both in an unbounded cell  \citep{Richardson1994,Richardson1996a,Crowdy2002,pof2012} and in the channel geometry \citep{Richardson1996b,CrowdyTanveer}. 
 \citet{Richardson2001} has obtained time-dependent solutions for injection flows in an unbounded cell  involving a multiply-connected fluid region, where the solution is given by a conformal mapping 
 written in terms of a ratio of Poincar\'e theta series associated with a Schottky group (the group $\Theta_0$ in our notation). 
 \citet{CrowdyMarshall2004}  used the formalism of the Schottky-Klein prime functions to construct conformal maps to multiply connected quadrature domains,  of which the Hele-Shaw flow domains considered by \citet{Richardson2001} are a subclass. 
  It would be interesting to investigate whether this approach could be generalised to other geometries.
Recently, time-dependent solutions have  also been  obtained  for injection flows  in an unbounded  Hele-Shaw cell in the presence of both a flat plate   \citep{Jonathan2015a} and a circular cylinder \citep{Jonathan2015b}, in which case the flow domain becomes doubly connected after the fluid has fully engulfed the obstacle.
Extension of these results to the case of multiple obstacles---thus implying multiply connected domains---should in principle be possible with the help of the secondary prime functions.

Particularly  relevant to the work presented herein is  the class of  time-dependent solutions for a single bubble in a Hele-Shaw channel recently obtained by \citet{MarkGio}. Their solution is a 
direct extension to the time-dependent case of the steady solution for a single bubble in a channel \citep{TS,TanveerBub}; and so it is natural to expect that the steady solutions reported here should  also lead to a  generalisation for  time-dependent flows with multiple bubbles. Constructing time-dependent solutions 
for moving bubbles in a Hele-Shaw cell 
is  of special  importance in connection with  
 the so-called selection problem, which concerns the question as to why a bubble or a finger advances with a unique velocity---twice that of the background fluid velocity---, despite the fact there is a continuum of steady solutions  for zero surface tension.  It was generally accepted that surface tension
was necessary for selecting the 
pattern \citep{Shraiman1986, HongLanger,Combescot, TanveerBub}.  However, recent studies \citep{Mineev1998, MarkGio, Robb2015}
have  demonstrated, by using
time-dependent exact solutions without surface tension, that
the steady solution with $U=2$ is the only attractor for the non-singular unsteady solutions, thus showing that velocity selection in a Hele-Shaw cell  is inherently determined by the zero surface tension dynamics. 
 It has  been conjectured \citep{MarkGio} that a similar scenario holds in domains of arbitrary connectivity.  The  steady solutions presented here---in terms of the secondary S-K prime functions---should  thus serve as a starting point to address  the  general selection problem for multiple bubbles. Work in this direction is currently underway. It is also hoped that the methods employed in the present work can be adapted to study other related physical systems, such as equilibrium configurations of multiple hollow vortices and the formation of multiple streamers in strong electric fields.

\begin{acknowledgments}
The author is  appreciative of the hospitality of the Department of Mathematics at Imperial College London (ICL) where this work was carried out. He  acknowledges financial support from a scholarship from the Conselho Nacional de Desenvolvimento Cientifico e Tecnologico (CNPq, Brazil) under the Science Without Borders program for a sabbatical stay at ICL. Helpful discussions with D.~Crowdy are gratefully acknowledged. The author would also like to thank J.~Marshall for a critical reading of the manuscript and for his help in generating the data used in figures 14 e 15.
\end{acknowledgments}


\bibliographystyle{jfm}

\end{document}